\documentclass[twocolumn]{aa}  

\usepackage{graphicx}
\usepackage{graphicx}
\usepackage{amssymb}
\usepackage{gensymb}
\usepackage{xspace}
\usepackage{ifthen}

\usepackage{txfonts}
\newcommand{\nnhp}{$\rm N_2H^+$\xspace}

\newcommand{\hhdp}{$\rm H_2D^+$\xspace}
\newcommand{\nndp}{$\rm N_2D^+$\xspace}
\newcommand{\ohhdp}{$\rm \text{o-} H_2D^+$\xspace}

\newcommand{\olineh}{$\rm \text{o-} H_2D^+(1_{1,0} - 1_{1,1})$\xspace}
\newcommand{\kms}{$\rm km \, s^{-1}$\xspace}
\newcommand{\ncol}{$N_\mathrm{col}$\xspace}
\newcommand{\xmol}{$X_\mathrm{mol} (\text{\ohhdp})$\xspace}
\newcommand{\tex}{$T_\mathrm{ex}$\xspace}
\newcommand{\vlsr}{$V_\mathrm{lsr}$\xspace}
\newcommand{\mcore}{$M_\mathrm{core}$\xspace}
\newcommand{\sigmav}{$\sigma_\mathrm{V}$\xspace}

\begin{document}

   \title{Identification of prestellar cores in high-mass star forming clumps via \hhdp observations with ALMA}


   \author{E. Redaelli
          \inst{1}
          \and
          S. Bovino \inst{2}
          \and
          A. Giannetti \inst{3}
          \and
          G. Sabatini \inst{2,3, 4}
          \and
          P. Caselli \inst{1}
          \and
          F. Wyrowski  \inst{5} 
          \and 
          D. R. G. Schleicher \inst{2}
          \and
          D. Colombo \inst{5}
          }

   \institute{Centre for Astrochemical Studies, Max-Planck-Institut f\"ur extraterrestrische Physik, Gie{\ss}enbachstra{\ss}e 1, 85749 Garching bei M\"unchen,
Germany\\
              \email{eredaelli@mpe.mpg.de}
         \and
            Departamento de Astronom\'ia, Facultad Ciencias F\'isicas y Matem\'aticas, Universidad de Concepci\'on, Av. Esteban Iturra s/n Barrio
Universitario, Casilla 160, Concepci\'on, Chile 
	\and
	Dipartimento di Fisica e Astronomia, Universit\`{a} degli Studi di Bologna, Via Gobetti 93/2, I-40129 Bologna, Italy 
	\and
	 INAF - Istituto di Radioastronomia - Italian node of the ALMA Regional Centre (It-ARC), Via Gobetti 101, I-40129 Bologna, Italy 
	 \and  Max-Planck-Institut f{\"u}r Radioastronomie, Auf dem H{\"u}gel 69, 53121 Bonn, Germany
             }

   \date{}

 
  \abstract
   {The different theoretical models concerning the formation of high-mass stars make distinct predictions regarding their progenitors, i.e. the high-mass prestellar cores. However, so far no conclusive observation of such objects has been made.}
   {We aim to study the very early stages of high-mass star formation in two infrared-dark, massive clumps. Our goal is to identify the core population that they harbour and to investigate their physical and chemical properties at high spatial resolution.}
   {We obtained ALMA Cycle 6 observations of continuum emission at $0.8\, \rm mm$ and of the ortho-\hhdp transition at $372\, \rm GHz$ towards the two clumps. We use the \textsc{scimes} algorithm to identify substructures (i.e. cores) in the position-position-velocity space, finding 16 cores. We model their observed spectra using a Bayesian fitting approach in the approximation of local thermodynamic equilibrium. We derive the centroid velocity, the linewidth, and the molecular column density maps. We also study the correlation between the continuum and molecular data, which in general do not present the same structure.}
   {We report for the first time the detection of ortho-\hhdp in high-mass star-forming regions performed with an interferometer. The molecular emission shows narrow and subsonic lines, suggesting that locally the temperature of the gas is less than $10\,\rm K$. From the continuum emission we estimate the cores' total masses, and compare them with the respective virial masses. We also compute the volume density values, which are found to be higher than $10^{6}\, \rm cm^{-3}$.}
   {Our data confirm that ortho-\hhdp is an ideal tracer of cold and dense gas. Interestingly, almost all the \hhdp-identified cores are less massive than  $\approx 13 \rm \,  M_\odot$, with the exception of one core in AG354, which could be as massive as $39\rm \, M_\odot$ in the assumption of low dust temperature ($5\, \rm K$). Furthermore, most of them are subvirial and larger than their Jeans masses. These results are hence difficult to explain in the context of the turbulent accretion models, which predict massive and virialised prestellar cores. We however cannot exclude that the cores are still in the process of accreting mass, and that magnetic fields are providing enough support for the virialisation. ALMA could also be seeing only the innermost parts of the cores, and hence the total cores' masses could be higher than inferred in this work. Furthermore, we note that the total masses of the investigated clumps are below the average for typical high-mass clumps, and thus studies of more massive sources are needed.}


\titlerunning{Prestellar cores in HM clumps identified with \hhdp}
   \maketitle
%

\section{Introduction\label{Introduction}}
It is generally known that stars form from the fragmentation and subsequent collapse of the cold molecular phase of the interstellar medium (ISM). However, while the star formation process in the low-mass regime is fairly well understood, we still lack a comprehensive view of how high-mass stars ($M> 8-10 \, \mathrm{M_\odot}$) are born. On the other hand, these play a key role in the energetic budget of the ISM. Furthermore, there is evidence that our Sun was born in a cluster containing also massive stars \citep{Adams10, Pfalzner20}. For all these reasons, the study of high-mass star formation is one of the crucial and still unanswered questions of modern astrophysics.  \par
In this context, several competing theories have been developed \citep{McKee03, Bonnell06, Smith09, Motte18, Padoan20}. Important differences among these models concern the very early stages of the process, characterised by the formation and evolution of so-called high-mass prestellar cores (HMPCs). In particular, the different theoretical models make distinct predictions on the masses, accretion modes, and in general on the dynamical and physical initial stages of HMPCs. Observations targeting high-mass star forming clumps, aimed to investigate the mass distribution and kinematic structure of such objects, could therefore provide crucial constraints to the theory. \par
Such observations are however difficult to perform, due to several reasons. High-mass stars are intrinsically rarer and more short-lived with respect to low-mass counterparts, as predicted by stellar evolution theories. As a consequence, they are on average more distant, which in turn affect the achievable spatial resolution of the observations. Moreover, they form in crowded and dense environments, heavily affected by extinction (e.g. \citealt{Zinnecker07}). This means that the lack of infrared emission detected with single-dish facilities is not a conclusive evidence of prestellar stage (e.g. \citealt{Motte18}). In this context, interferometers such as  the Atacama Large Millimeter/submillimeter Array (ALMA) represent a powerful tool, being able to provide the necessary angular resolution and sensitivity to resolve substructures in distant high-mass clumps. \par
In the context of the search for HMPCs, targeting $70 \mu \rm m $-dark clumps with interferometric studies has been a powerful tool \citep{Sanhueza17, Contreras18, Li19, Pillai19}.  In \cite{Molet19}, the authors investigate with ALMA band 6 observations two HMPCs candidates in the rich high-mass star forming regions W43-MM1, previously investigated by \cite{Nony18}. One of the two cores does not show clear evidence of protostellar activity, but even the ALMA resolution is not sufficient to draw clear conclusions on its nature. Using  Submillimeter Array (SMA) and Very Large Array (VLA) data, \cite{Cyganowski14} studied the massive starless core G11.92-0.61-MM2, and found that, despite its high density ($n \gtrsim 10^9\, \rm cm^{-3}$) and mass ($M \gtrsim 30 \, \rm M_\odot$), it lacks line emission in several, abundant species such as \nnhp, $\rm HCO^+$, $\rm HCN$ (down to the sensitivity of their observations). \cite{Pillai19} investigated two infrared dark clouds (IRDCs), selected to be $70\, \mu \rm m$-dark and therefore believed to be prestellar, with the SMA. The authors found several substructures in the dust thermal emission data. However, using CO (2-1) observations, they found that both IRDCs host a large population of molecular outflows, indicating that several protostars are already active. The ASHES survey (Alma Survey of $70
\,\mu \rm m$ dark High-mass clumps in Early Stages; \citealt{Sanhueza19, Li20}) targeted twelve IRDCs dark in the wavelengths  $3.6-70\, \mu \rm m$ with ALMA at $1.3\, \rm mm$ continuum and with several molecular tracers. They identified $\approx 300$ cores, and based on the detection of outflows and/or of warm transitions\footnote{Sanhueza et al. (2019) define "warm transitions" those with upper level energies higher than $22 \, \rm K$.} $\approx 70$\% of them are classified as prestellar. None of them appears more massive than $10-30\, \mathrm{M_\odot}$. \par
These examples highlight the difficulties not only in observing cores in the high-mass regime, but also in correctly classifying them as pre-/proto-stellar. Continuum observations at mm/sub-mm wavelengths (especially when only one frequency is available) cannot provide information on the mass and on the temperature independently, and they cannot unveil the sources kinematics either. Other wavelengths could provide useful information in this sense, as done for instance with X-ray observations in \cite{Yu20}.
On the other hand, chemistry is a powerful tool to probe the evolution of star forming regions. In the cold molecular phase of the ISM, at low temperatures ($T \lesssim 20 \, \rm K$) and high densities ($n \gtrsim 10^4 \, \rm cm^{-3}$), most C- and O-bearing species are frozen out onto dust grains \citep{Caselli99, Bacmann02, Giannetti14, Sabatini19} and they are therefore depleted from the gas phase. In these physical conditions processes such as deuteration are greatly enhanced \citep{Caselli12,Ceccarelli14}. In particular, the first deuterated species in the gas phase is produced by the reaction:
\begin{equation}
\label{Deuteration}
\mathrm{H_3^ + + HD} \leftrightharpoons \mathrm{H_2D^+ +H_2} + 230\text{K}\, ,
\end{equation}
which is particularly efficient at low temperatures (assuming a low ortho-to-para $\rm H_2$ ratio; see e.g. \citealt{Pagani92}), due to the lower zero-point energy of \hhdp. From this, and from its doubly and triply deuterated forms $\rm D_2 H^+$ and $\rm D_3^+$, all other deuterated molecules in the gas phase (such as $\rm N_2D ^+$ and $\rm DCO^+$) are formed \citep{Dalgarno84}.  \par
In a hunt for HMPCs, Tan and collaborators have used an extensive dataset of \nndp observations with ALMA. \cite{Tan13} reported initially the detection of six cores, the most massive of which was revealed to be in fact composed by several cores, some of which already in a protostellar stage \citep{Tan16, Kong18}. The sample was increased to 141 \nndp-identified cores in \cite{Kong17}. Interestingly, only one core appears to be more massive than $10 \, \rm M_\odot$. However, the lack of tracers of protostellar activity leaves the question about the evolutionary stage of this core still open.  \par
Deuteration of simple molecules has indeed raised huge interest recently, since it has been shown to be a useful ``chemical clock'', correlated with the evolutionary stage of star forming regions, as shown by \cite{Emprechtinger09} in the low-mass regime. \cite{Fontani11} showed that the $\rm N_2D^+/N_2H^+$ isotopic ratio decreases from $\approx 0.2-0.4$ to less than $0.1$ when the temperature rises above $20\, \rm K$ in a sample of high-mass star forming regions in different evolutionary stages. \cite{Feng19,Feng20} reported similar findings using $\rm N_2H^+$ and $\rm HCO^+$ isotopologues.  Despite the fact that \nndp is an abundant ion in cold and dense gas, it may not be the ideal tracer of HMPCs. N-bearing species are less affected by freeze-out onto the dust grain surfaces than molecule containing carbon and/or oxygen, but at high densities ($n \gtrsim 10^{6} \, \rm cm^{-3}$) \nnhp and \nndp show sign of depletion as well \citep{Pagani07,Redaelli19}. \par
Transitions of lighter, deuterated molecules with high critical densities then represent the only reliable tracers of prestellar gas, since they can persist in the gas phase at high densities. \hhdp is an ideal candidate, as shown by the complete depletion model of \cite{Walmsley04}. The $(1_{1,0} - 1_{1,1})$ transition of its ortho form (hereafter \ohhdp) has a critical density of $n_\mathrm{cr} \approx 10^5 \, \rm cm^{-3}$ \citep{Hugo09}. Furthermore, \hhdp is very sensitive to the temperature, due to the chemical reactions with its main destroyer, carbon monoxide. When ---due for instance to protostellar activity--- the gas temperature raises beyond $20 \, \rm K$, the CO molecules frozen onto the dust grains evaporate back into the gas phase, thus lowering the \hhdp abundance. Moreover, above $30\, \rm K$ reaction \eqref{Deuteration} start to proceed backwards, thus reducing the deuteration level. Hence, detecting \hhdp is an unambiguous sign of prestellar conditions in the sense that core structures which show significant emission in \ohhdp are trustfully in a prestellar phase.  \par

\begin{figure*}[!h]
\centering
\includegraphics[width=1\textwidth]{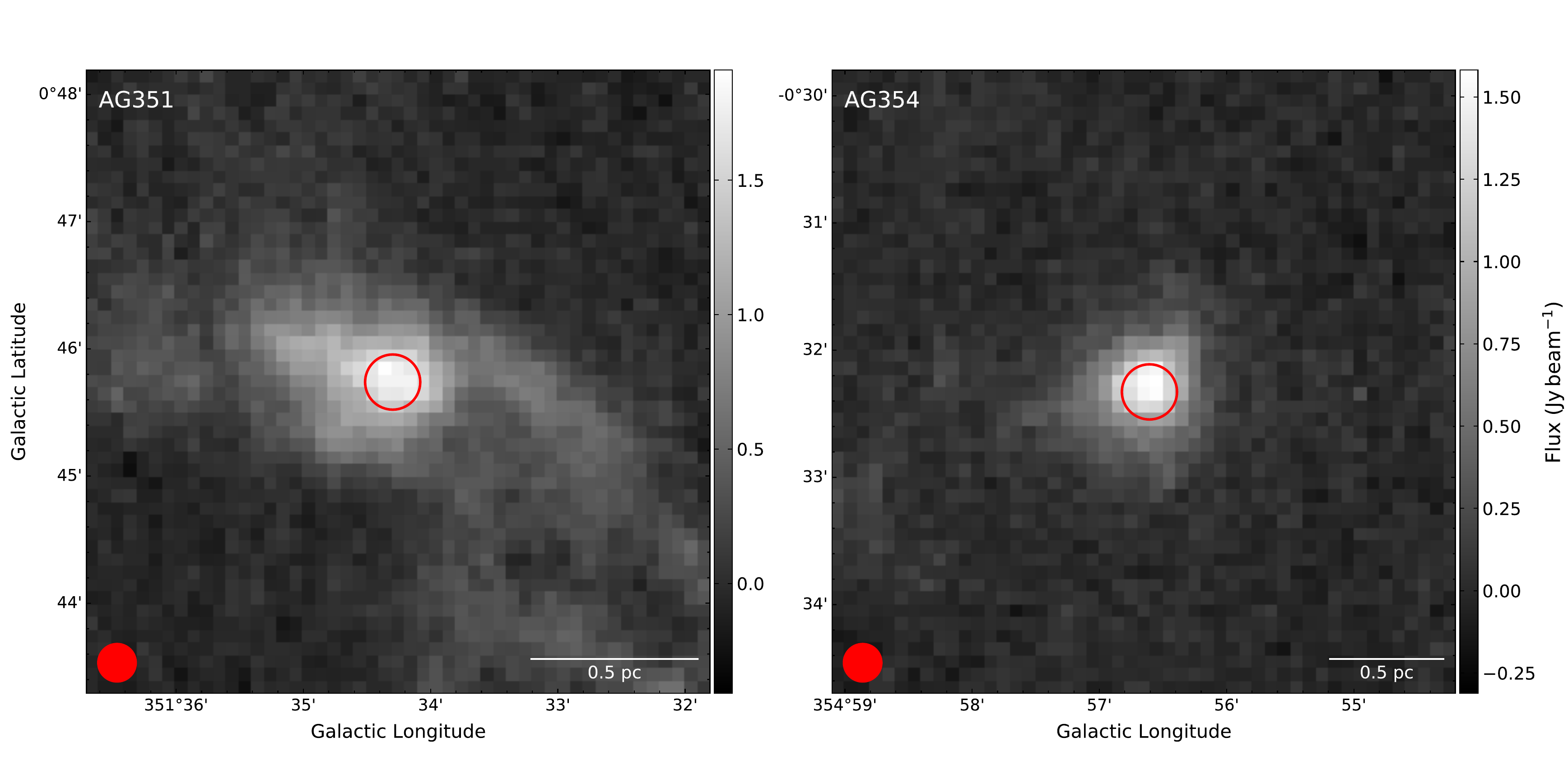}
\caption{Maps of the continuum emission at $870 \, \rm \mu m$ seen with APEX by the ATLASGAL survey in AG351 (left) and AG354 (right). The FoV of the ALMA observations is indicated with a red circle. The beam size ($18.2''$) is shown in the bottom-left corner, whilst the scalebar is in the bottom-right.\label{LABOCA}}
\end{figure*}

Studies of \hhdp in star forming regions are however rare, and mainly performed with single-dish facilities, due to the weakness of its transitions. \cite{Caselli03,Caselli08} produced a small sample of low-mass cores observed in \olineh, and \cite{Miettinen20} targeted dense cores in Orion B9. In the high-mass regime, \cite{Pillai12} spatially resolved the \hhdp emission in the star forming region DR21, in Cignus X, using JCMT. \cite{Bruenken14} demonstrated the use of \hhdp as a chemical clock for low-mass star forming regions. More recently,  \cite{Giannetti19} detected \olineh in three clumps belonging to the same filament with the Atacama Pathfinder EXperiment telescope \citep[APEX;][]{Gusten06}, showing that the abundance of this molecule anticorrelates with that of \nndp. In a following study, \cite{Sabatini20} published a detailed census of this tracer in 16 high-mass clumps with the APEX telescope. To our knowledge, only one interferometric study of this molecule has been performed: \cite{Friesen14} used ALMA band 7 observations to target the low-mass star forming region Ophiuchus. No high-mass dedicated study of \hhdp with interferometers is present in literature. \par
In this work, we report the first interferometric observation of \olineh in two high-mass clumps performed with ALMA at $\approx 1''$ resolution. The line was successfully detected in both sources, and its emission appears extended. Using a core-identifying algorithm, we find a total of 16 cores in \hhdp. We study their properties using a spectral fitting analysis. Interestingly, a significant fraction of the gas emitting in \olineh presents low, subsonic velocity dispersion, often lower than the thermal broadening of the line at $10\, \rm K$. This indicates that part of the clumps is still in a very cold, prestellar stage. We study the correlation of the molecular emission and the dust thermal emission, also detected with ALMA at $0.8 \rm \, mm$. The line integrated intensity usually does not correlate with the continuum data, and we speculate that this is due to different evolutionary stages of the identified cores. Among them, two present both bright \ohhdp emission and continuum emission. They represent ideal candidates to be prestellar cores embedded in high-mass clumps. \par
The observations and the resulting data are described in Sects. \ref{Observations} and \ref{Results}. We present the data analysis in Sect. \ref{Analysis}, where we identify cores in position-position-velocity space in the \ohhdp data-cubes and we perform the line spectral analysis. We discuss the results in Sect. \ref{Discussion}: we first focus on the properties of the gas traced by the \ohhdp line (Sect. \ref{Disc:1}), and then on the comparison between the molecular and the continuum data (Sect. \ref{Disc:2}), using the dust thermal emission to estimate the gas column density and the sources' total masses. In Sect. \ref{Disc:3} we speculate on the possible evolutionary sequence of the identified cores. We discuss our findings in the context of star formation theories in Sect. \ref{Disc:4}. Section \ref{Summary} contains the summary 
and concluding remarks of this work.

\section{Source selection and observations\label{Observations}}

\subsection{The targeted sources}
The sources selected for this work belong to the APEX Telescope Large Area Survey of the Galaxy (ATLASGAL), which comprises a large number ($\sim 10000$) of massive clumps in different evolutionary stages \citep{Schuller09}. In particular, the two chosen clumps belong also to the ATLASGAL TOP100 sub-sample \citep{Giannetti14,Konig17}, and their ID are AGAL351.571+00.762 (hereafter AG351) and AGAL354.944-00.537 (hereafter AG354). Figure \ref{LABOCA} shows the dust thermal emission at $870\, \rm \mu m$ as seen with APEX at a resolution of $\approx 18''$. The clumps have similar masses and distances: $M_\mathrm{clump} = 170 \rm \, M_\odot$ and $D = 1.3 \,  \rm kpc$ for AG351,  $M_\mathrm{clump} = 150 \rm \, M_\odot$ and $D = 1.9 \,  \rm kpc$ for AG354 (see \citealt{Konig17} and references therein). It is worth noting that these clumps have masses below the median value of the distribution ($\sim500 \mathrm{M_\odot}$) obtained for the quiescent stage classified in the ATLASGAL sample \citep[see][]{Urquhart18}. \cite{Konig17} investigated the dust emission properties in the TOP100 sample using continuum data in the wavelength range $3-870 \, \mu \rm m$ from several infrared surveys. From the fit of the spectral energy distribution they estimated dust temperatures of $17-19\, \rm K$, which must be considered average values over the source sizes ($80-100''$). \par
Concerning their evolutionary stage, both AG351 and AG354 are classified by the ATLASGAL catalogue as $24\,\rm \mu m$-dark, since they are not associated to detected point-like sources at $24$ and $70\, \rm \mu m$. We have furthermore inspected the available archive data at mid-infrared wavelengths: the Herschel-PACS maps at $70\, \rm \mu m$ and the Spitzer-IRAC maps at $3.6, 4.5, 5.8, \text{and } 8.0\, \rm \mu m$ (published in the GLIMPSE survey; \citealt{Benjamin03, Churchwell09}). No point source is detected within a radius of $\approx 30''$. In addition, \cite{Kuhn20} produced a Spitzer/IRAC catalogue of protostellar candidates in the Galactic midplane, and they did not find young stellar objects associated with the positions of the investigated clumps. This supports the scenario according to which AG351 and AG354 are in early prestellar phases, even though we highlight that several works showed that 70$\mu$m-dark clumps were later discovered to contain protostars with intereferometric observations \citep{Pillai19, Li19, Li20}.\par
Both clumps are included in the sample investigated by \cite{Sabatini20}, who detected the \olineh transition with APEX in 16 ATLASGAL clumps. Their data have a beam size of $16.8''$ and spectral resolution of $0.55\,$\kms. In AG351 and AG354 the line peak temperature is $T_\mathrm{peak}\approx 200\, \rm mK$ and the linewidth is $ \approx 1.0-1.2 \,$\kms. Using the $\rm H_2$ column densities from literature data \citep{Konig17, Urquhart18}, \cite{Sabatini20} estimated an abundance with respect to molecular hydrogen of $X_\mathrm{mol}(\text{\ohhdp}) \approx 10^{-10}$ in both clumps.

\begin{figure*}[!t]
\centering
\includegraphics[width=1\textwidth]{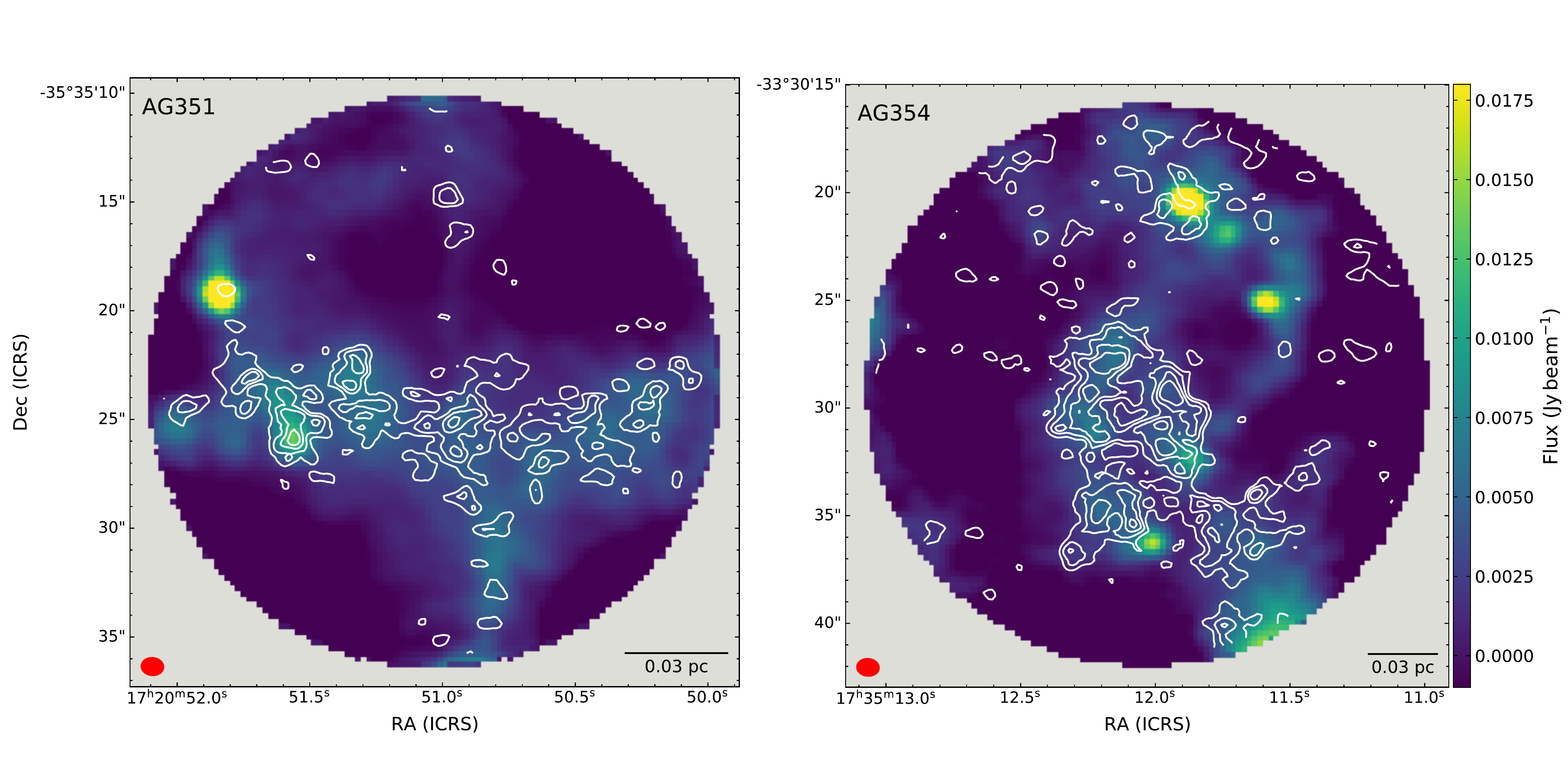}
\caption{Map of the continuum emission as seen by ALMA in band 7 in AG351 (left panel) and AG354 (right panel). The contours show the \ohhdp integrated intensity emission from the original not regridded, and not primary-beam corrected data, at levels $=[5,7,9,11]\sigma$, where $1\sigma=60 \, \rm mK\,$\kms (AG351) and $70 \, \rm mK\,$\kms (AG354). For the continuum data, we show the primary-beam corrected maps. The beam size and the scale bar are shown in the bottom-left and bottom-right corners, respectively. \label{ResultsFig}}
\end{figure*}

\subsection{Observations and data reduction \label{Obs}}

AG351 and AG354 were observed during Cycle 6 as part of the ALMA project \#2018.1.00331.S (PI: Bovino) in two runs, from November 2018 to April 2019. The observations made use of both the Main Array (12m-array 45 antennas) and the Atacama Compact Array (ACA, 12 antennas), with baselines ranging from $9$ to $313 \, \rm m$. They were acquired as single-pointings, centred at the sources' coordinates: $\text{RA}= 17^h20^m51.0^s$, $\text{Dec}=-35^{\circ}35'23.29''$ for AG351 and $\text{RA}=17^h35^m12.0^s$, $\text{Dec} =-33^{\circ}30'28.97''$ for AG354 (J2000). The quasars J1700-2610, J1717-3342, J1924-2914, and J1517-2422 were used as calibrators. \par
The spectral setup consists of four spectral windows (SPWs). One is centred on the \ohhdp $(1_{1,0} - 1_{1,1})$ line at the frequency $\nu_\mathrm{rest} = 372421.3558\, \rm MHz$ \citep{Jusko17}. This SPW has a resolution of $244 \, \rm kHz$ (corresponding to $0.20\,$\kms at  $372 \, \rm GHz$) and a total bandwidth of $500 \, \rm MHz$. A second spectral window is dedicated to the continuum, with a total bandwidth of $2.0 \, \rm GHz$, centred at the frequency of $371\, \rm GHz$. The remaining two SPWs were centred on $\rm SO_2$ and methanol lines  {, with a total bandwidth of 938$\,$MHz and spectral resolution of 564$\,$kHz ($\approx 0.47\,$\kms). These lines however} were not detected. At these frequencies, and with the used configuration, the maximum recoverable scale is $\theta_\mathrm{MRS} \approx 20''$, the primary beam size is $\theta_\mathrm{FoV} \approx 26''$\footnote{At $372\, \rm GHz$, the primary beams of the main array and of ACA are $17''$ and $30''$, respectively}, and the resolution is $\approx 1.0''$ (corresponding to $\approx 1600 \, \rm AU$ at the sources average distance of $1.6\, \rm kpc$).  The total observing time were $180\, \rm min$  (ACA) and $28\, \rm min$ (12m-array) for each source. During the observations, the precipitable water vapour was typically $0.3 \, \mathrm{mm}< PWV < 0.7\, \mathrm{mm}$, and in general lower  than $1\,\rm mm$. The average system temperature values are found in the range $300-400\, \rm K$ for the SPW containing the \olineh line. \par
 The data were calibrated by the standard pipeline (\textsc{casa}, version 5.4; \citealt{McMullin07}). The quality assessment outputs were checked, and the automatic calibration results are satisfactory.  From a first inspection of the dirty maps, the emission both in continuum and in line appear very extended in the whole Field-of-View (FoV). We therefore applied a modified weight of $2.4$ to the ACA observations, chosen to maximise the recovery of large-scale flux, without downgrading the resolution. The ACA and 12m-array observations were then concatenated to proceed with the imaging. \par 
Both continuum and line data were imaged with the \textsc{casa} \texttt{tclean} task interactively (\textsc{casa} version 5.6), using a natural weighting and the multiscale deconvolver algorithm \citep{Cornwell08}. In order to maximise the bandwidth used to produce the continuum maps, we combined the continuum-dedicated SPW with the line-free channels in the other windows, obtaining a total bandwidth of $\approx 4.2 \, \rm GHz$. In order to obtain the data-cubes of the \ohhdp line, we focused on the central $25\, \rm MHz$ ($\approx 20 \, \rm km \, s^{-1}$) around the source centroid velocity. The data-cubes hence present 100 channels in the frequency axis, and the final spectral resolution is $0.22\,$\kms. Finally, in order to avoid oversampling, both the continuum and the line images (already primary-beam corrected) have been re-gridded in order to ensure 3 pixel per beam minor axis, in agreement with the Nyquist theorem. The molecular line data have been converted into the brightness temperature $T_\mathrm{b}$ scale, computing the corresponding gain, $G= 11 \, \rm mK/ (mJy\, beam^{-1})$ at a beam size of $1''\times0.8''$. In Appendix \ref{APEX} we discuss the possible missing flux problem affecting our data, by comparing single-dish and interferometric observations.

\section{Results\label{Results}}
Figure \ref{ResultsFig} shows the continuum maps obtained as described in Sect. \ref{Obs} for AG351 and AG354, respectively. The contours show the integrated intensity maps of the \olineh transition, obtained integrating the cubes over the range $[-4.3; -2.0]$ \kms (AG351) and $[-7.5; -4.0]$ \kms (AG354). The continuum maps present a sensitivity of $0.8\, \rm mJy/beam$ for both sources, whilst the median root-mean-squared ($rms$) of the line data, computed over emission-free channels, is $ 300 \, \rm mK$ per $ {0.22\,}$\kms. These values refer to already primary-beam corrected data. The observation resolutions and sensitivities for each source are summarised in Table \ref{ObsSummary}.

\begin{table*}[]
\renewcommand{\arraystretch}{1.4}
\centering
\caption{Beam sizes, spatial resolutions, achieved sensitivity ($rms$), and spectral resolutions of the continuum and line observation in AG351 and AG354. The $rms$ values have been computed on the primary-beam corrected data.}
\label{ObsSummary}
\begin{tabular}{ccccc}
\hline
           Observation                   & Beam size\tablefootmark{a}            & Spatial res.         & $rms$                  & Spectral res.        \\
 
                              \hline  \hline  
                               \multicolumn{5}{c}{AG351}                                                              \\
Continuum                     &       $1''.03\times 0''.82$, $PA=84.7$\degree     &  $1340 \rm AU \times 1170 AU$  &      $0.8 \, \rm mJy/beam$       &        -              \\
\olineh                         &     $1''.00\times 0''.80$, $PA=85.9$\degree          &     $1300 \rm AU \times 1040 AU$   &      $300 \, \rm mK$ &   $0.22\,$\kms        \\
\hline
                               \multicolumn{5}{c}{AG354}                                                              \\
Continuum &    $1''.04\times 0''.82$, $PA=84.7$\degree  &  $1980 \rm AU \times 1560 AU$ &$0.8 \, \rm mJy/beam$ & -  \\
\olineh                         &   $1''.01\times 0''.79$, $PA=85.7$\degree    &     $1920 \rm AU \times 1500 AU$     &      $300 \, \rm mK$     &       $0.22\,$\kms             \\ 
\hline
\end{tabular}
\tablefoot{
\tablefoottext{a}{The beam size is expressed as: major axis $\times$ minor axis, position angle ($PA$).}}
\end{table*}

\section{Analysis \label{Analysis}}
Our goal is to identify substructures seen in the \ohhdp emission in each clump, and to analyse their physical and chemical properties. To reach our aim we have made use of an automated algorithm that allows to identify such structures, to fit the molecular line spectra, and to reconstruct properties such as their column density,  velocity dispersion, and centroid velocity. In the following subsections we describe the steps of our analysis.

\subsection{Core identification with \textsc{scimes} \label{SCIMES}}
In order to identify substructures in the molecular emission we use the Spectral Clustering for Interstellar Molecular Emission Segmentation package (\textsc{scimes}, \citealt{Colombo15}). \textsc{scimes} is based on dendrograms algorithms \citep{Rosolowski08}, and it is developed specifically to analyse molecular emission data in the form of position-position-velocity cubes. It identifies the substructures at the basis of the dendrogram (\textit{leaves}), and it then finds the parental structures to which they belong (\textit{branches} and \textit{trunks}). To our scopes, we are interested in the smallest, significant substructures visible in \ohhdp emission. We therefore focus on the leaves, which in the context of this paper correspond to prestellar cores. \par
In order to first build the dendrogram, a number of parameters is needed. We performed multiple tests, followed by visual inspection of the results, to find the best choice for our \ohhdp data. First of all, following \cite{Rosolowsky06}, we find the regions in the emission characterised by signal-to-noise ratio ($\rm S/N$) higher than a given threshold ($\rm S/N_{lim}$), which contain peaks of emission brighter than a second higher threshold ($\rm S/N_{peak}$). This implementation maximises the recovered information in case of moderate-to-low signal-to-noise-ratio data ($\rm S/N = 3-10$), such as our \ohhdp ALMA observations. In fact, the lines are locally bright ($T_\mathrm{b}>0.5\, \rm K$), but the peak temperature rapidly drops. We therefore set $\rm S/N_{peak} = 2$, and $\rm S/N_{lim} = 1.5$ for AG351, and $\rm S/N_{peak} = 2$, and $\rm S/N_{lim} = 1.3$ for AG354. By doing so, we set the minimum value ($min_\mathrm{val}$) of the dendrogram algorithm to zero. \par 
Another key parameter to build the dendrogram is the minimum height (in flux/brightness) that a structure must have to be catalogued as an independent leaf ($\Delta_{min}$). In our case, we set $\Delta_{min} = 2.0 \times rms$. \textsc{scimes} performs better with data-cubes with constant noise, since $\Delta_{min}$ has to be set as a single value. We therefore apply the algorithm to the cubes before applying the primary-beam correction. Their average noise is hence $rms = 150\, \rm mK$. We highlight that this correction applies a corrective factor dependent on the position within the primary beam, but uniform in the frequency axis. Therefore, it alters the absolute value of the $rms$, which becomes larger at the edges of the primary beam, but it does not affect the signal-to-noise ratio map. We set the minimum number of channels that a leaf must span to $N \rm^{min} _{chan} = 2$, due to the low linewidth of the \ohhdp line. Finally, we exclude leaves smaller than two times the beam area. All the identified cores are found within regions where the integrated intensity is detected above a $3\sigma$ level ($\rm S/N_\mathrm{II}>3.0$). 
\par
\begin{figure*}[h]
\centering
\includegraphics[width=\textwidth]{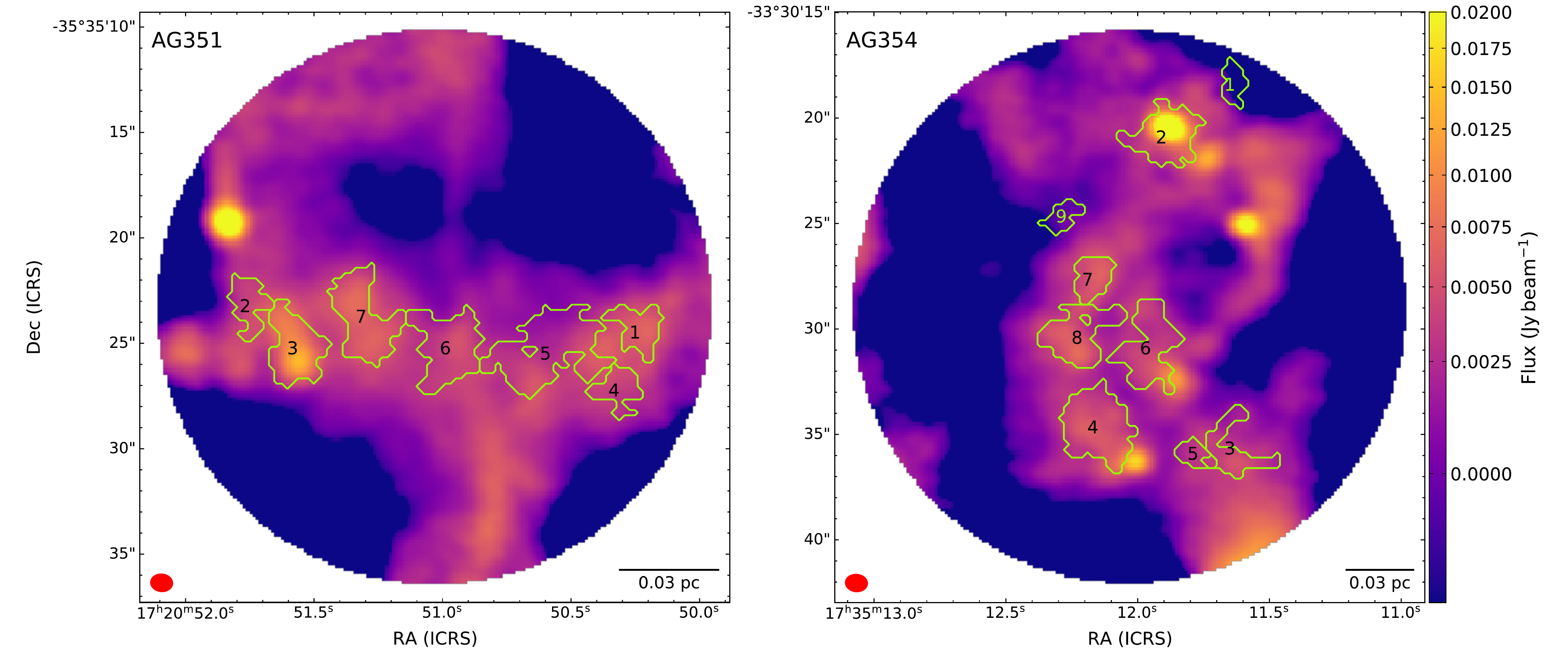}
\caption{The contours show the leaves identified by \textsc{scimes}, overlaid to the primary-beam corrected continuum map in AG351 (left panel) and AG354 (right panel). \label{Fig:leaves}}
\end{figure*}

At the end of the process we identify seven cores in AG351 and nine cores in AG354, which we have labelled in order of decreasing peak value of \ohhdp integrated intensity within each clump. Figure \ref{Fig:leaves} shows the identified structures on top of the continuum maps. It is worth noting that there are rare cases of cores overlap (e.g. cores 3 and 5 in AG354). This is due to the fact that \textsc{scimes} works in position-position-velocity space, and it is therefore able to disentangle structures which overlap spatially but present different centroid velocities (i.e. distinct velocity components along the line of sight). The first four columns of Table \ref{CoreProp} report the cores identification labels, centre positions, and effective radii ($R_\mathrm{eff}$, defined as the radius of a circle with the same area as the core). Several cores appear irregular in shape. This is due to \textit{i)} the limited spatial resolution of our observations, combined with a Nyquist sampling of the beam size, which hence tends to make the borders of the identified structures more irregular; \textit{ii)} the limited S/N of the \ohhdp data, which translates in the fact that the identified cores, despite being significant at a $3\sigma$ level in integrated intensity, are often at the limit of detection in peak brightness temperature; and \textit{iii)} the lack of total-power observations, which should recover the most extended emission surrounding the cores. Moreover, we highlight that even in simulations cores do not always appear regular in shapes (see e.g. \citealt{Smullen20}).

\par
Figure \ref{ResultsFig} shows that the \ohhdp integrated intensity distribution does not always follow the structure seen in continuum emission. The cores just identified confirm this scenario: in some cases, dust thermal emission peaks are found within molecular-identified cores (e.g. core 3 in AG351, or core 2 in AG354). In other cases, bright continuum spots do not overlap with \hhdp emission. In Sec. \ref{Disc:2} and Appendix \ref{App:contCores} we further investigate this point.

\begin{figure*}[h]
\centering
\includegraphics[width=1\textwidth]{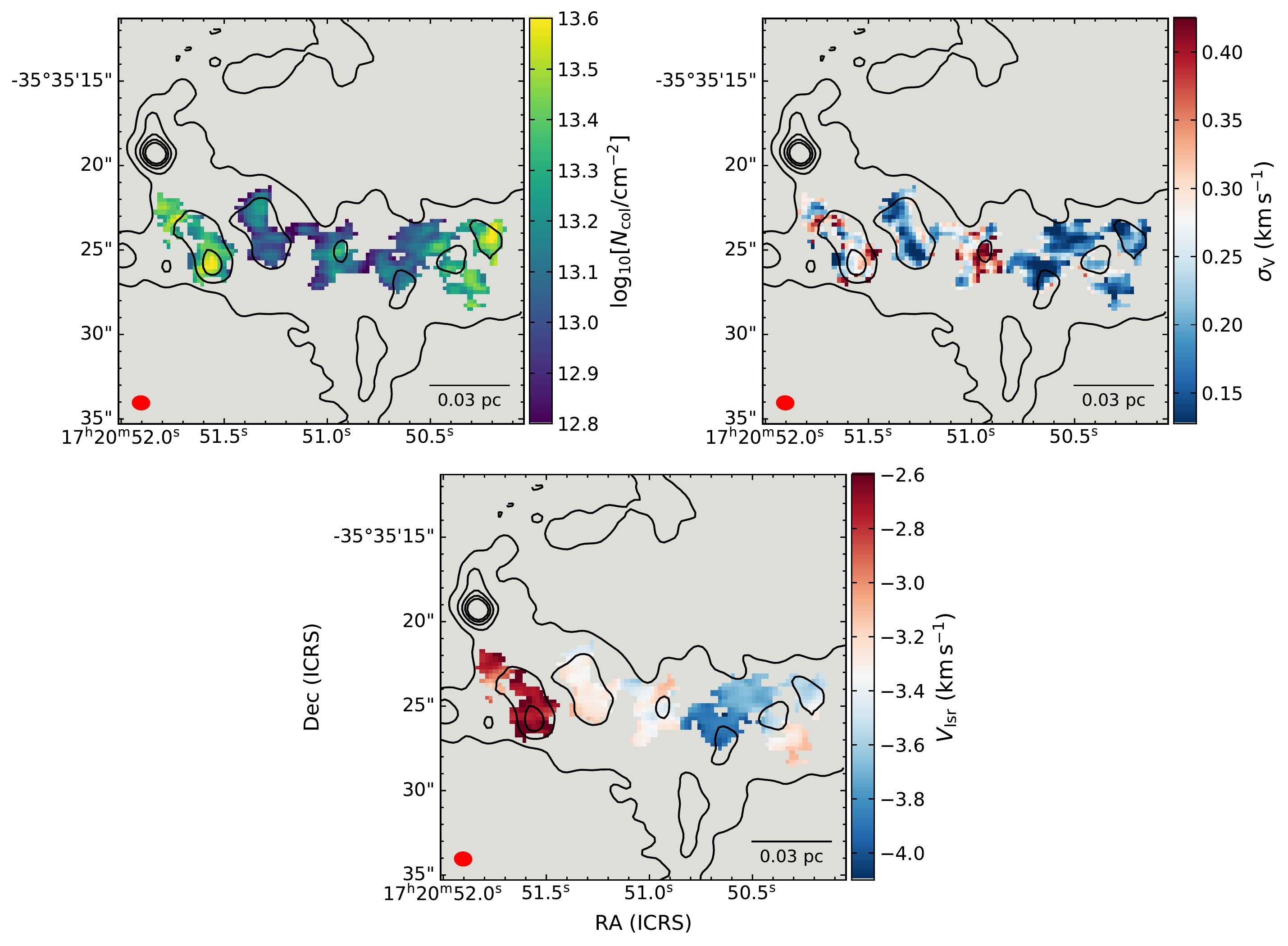}
\caption{Maps of \ohhdp column density (top left panel), \sigmav (top right), and centroid velocity (bottom) obtained from the \textsc{MCWeeds} analysis in the individual cores in AG351. The beam size and scalebar are shown in the bottom-left and bottom-right corners, respectively. The black contours show the continuum emission in band 7 at levels $[1, 5, 10, 15, 20]\, \rm mJy \,beam^{-1}$.\label{AG351_Maps}}
\end{figure*}

\begin{figure*}[h]
\centering
\includegraphics[width=1\textwidth]{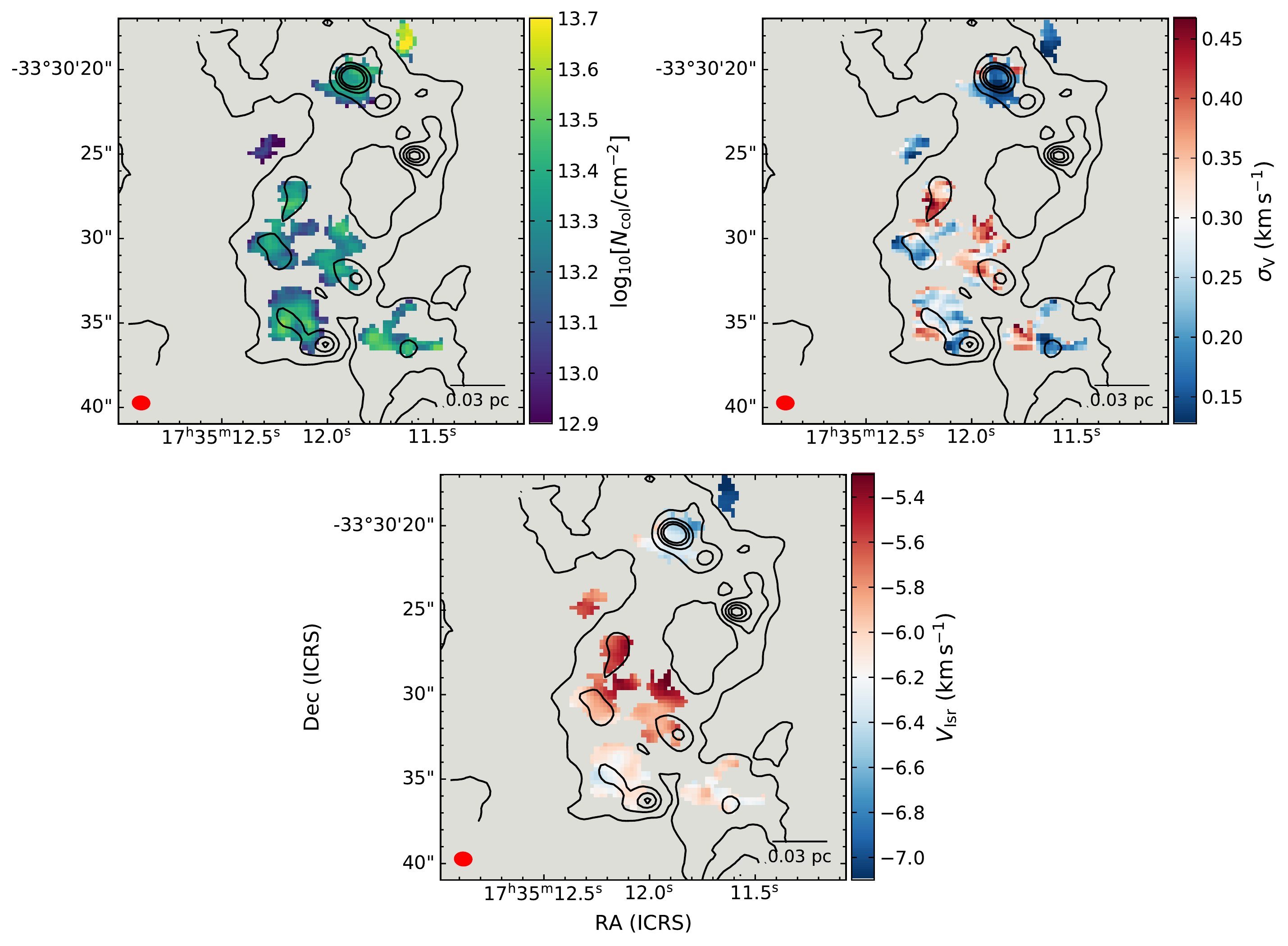}
\caption{Same as Fig. \ref{AG351_Maps}, but for AG354. \label{AG354_Maps} }
\end{figure*}
\subsection{Spectral fitting with \textsc{MCWeeds} \label{MCWeeds}}
In order to derive the physical parameters of each core from the \ohhdp data, we perform a spectral fitting of the lines pixel-per-pixel using \textsc{MCWeeds} \citep{Giannetti17}. Briefly, the code combines the \textsc{Weeds} package from \textsc{GILDAS} \citep{Maret11}, which is optimised to produce synthetic spectra assuming local thermodynamic equilibrium conditions (LTE), with Bayesian statistical models implemented using \textsc{PyMC} \citep{Patil10} to perform the actual fit. We use a Markov chain Monte Carlo (MCMC) algorithm to sample the parameter space, using uninformative flat priors over the models' free parameters. We recently upgraded \textsc{MCWeeds}, parallelising the code using the methods of the python package \textsc{multiprocessing}\footnote{\url{https://docs.python.org/3/library/multiprocessing.html}}. This improves the code speed when dealing with large datasets, allowing to fit simultaneously spectra from multiple positions. For each spectrum, the code performs 100000 iterations, and excludes the first 1000 steps (burn-in), which represents a reliable choice, as previously investigated \citep[see Appendix B in][]{Sabatini20}. The stability and convergence of the chains was in any case checked. \par
Since our ALMA data comprise only one \ohhdp transition, they do not allow to constrain independently the molecular column density (\ncol) and the excitation temperature (\tex). We therefore assume $\text{\tex} = 10 \, \rm K$, a choice consistent with previous works: \cite{Caselli08} found $\text{\tex} = 7-14 \, \rm  K$ in a sample of low-mass prestellar and protostellar cores; \cite{Friesen14} adopted $\text{\tex} = 12 \, \rm K$ when studying prestellar cores in Ophiuchus A. The remaining free parameters in the fit are hence the molecular column density $N\rm _{col}(\text{\ohhdp})$, the line local-standard-of-rest velocity $V_\mathrm{lsr}$, and the line full-width at half-maximum $FWHM$. For each core, individual initial guesses for the free parameters are chosen to improve the code convergence. In order to estimate correctly the parameters distributions, the noise level of the spectral line data has to be indicated. We adopt $rms = 180 \, \rm mK$ (AG351) and $rms = 170 \, \rm mK$ (AG354)\footnote{The reader may notice that these noise levels are different from those indicated in Table \ref{ObsSummary}. This is due to the fact that the $rms$ values in Sect. \ref{Results} are computed over the whole map, whilst in Sect. \ref{MCWeeds} we evaluate them only over positions belonging to cores identified by \textsc{scimes}. Since the map edges, which are noisier, do not present many cores, the $rms$ is reduced.}. \par
The parallelised version of \textsc{MCWeeds} returns the maps of the best fit parameters, together with those of the lower and upper 95\% high probability density (HPD) intervals. The best-fit models have been visually inspected, in order to assess their quality. In both clumps double-peak features, either due to multiple component on the line of sight, to self-absorption, or to large scale emission filtering are present in the observed \ohhdp spectra. Even though they represent a small minority of the observations, these spectra cannot be successfully fitted with a single velocity-component, LTE model. Since these positions are characterised by an overestimation of the linewidth, we find that a successful masking strategy is to discard fits with $FWHM$ higher than a given threshold. This threshold, which is chosen for each core individually, is in the range $1-1.2\,$\kms. The fraction of rejected pixels is $\lesssim 3$\% for both clumps, and we checked that no single-component pixels were masked. In Figs. \ref{AG351_Maps} and \ref{AG354_Maps} we show the resulting best-fit parameter maps in AG351 and AG354, respectively. Individual maps for each core are presented in Appendix \ref{App:allcores}. As mentioned, \textsc{MCWeeds} use the line $FWHM$ as free parameter. We however convert the obtained values of this quantity into velocity dispersion \sigmav, in order to allow a more straightforward comparison with, for instance, the isothermal sound speed. \par
The median values for the relative errors derived from the 95\% HPD intervals obtained with \textsc{MCWeeds} are: 57\% in $\sigma_\mathrm{V}$, 43\% in \ncol, and 3.5\% in \vlsr for AG351; 38\% in $\sigma_\mathrm{V}$, 33\% in \ncol, and 1.9\% in \vlsr for AG354. Uncertainties for AG351 are on average larger, since the spectra in this source are less bright. The centroid velocity is usually  well constrained, whilst the velocity dispersion is characterised by the highest uncertainties. These are particularly large at low $\sigma_\mathrm{V}$ values, due to the limited spectral resolution of our observations, which is discussed further later on. 
\par
The \vlsr maps in Fig. \ref{AG351_Maps} and \ref{AG354_Maps} show that the cores are coherent structures in velocity, with dispersions around  mean values of usually $\approx 0.2\,$\kms. At the clump scale, stronger velocity gradients of $\approx 1\,$\kms and $\approx 2\,$\kms are seen in AG351 and AG354, respectively. We avoid detailed discussion on the centroid velocity gradients seen in the clumps, since the lack of zero-spacing observations prevents us to infer the large-scale gas kinematics (see Appendix \ref{APEX} for more details). \par

Typical values for the molecular column density are in the range $N_\mathrm{col} = (0.6 - 4) \times 10^{13} \, \rm cm^{-2}$. These are higher than those found by \cite{Sabatini20} using APEX observations, most likely due the dilution of the \ohhdp emission in the large single-dish beam ($16.8''$). The observed values are on average higher also than the maximum column density reported by \cite{Friesen14} with ALMA at a higher spatial resolution ($N_\mathrm{col} = 1.2 \times 10^{13} \, \rm cm^{-3}$ at $\sim 150 \, \rm AU$). However, those authors targeted a low-mass star forming region, focusing on a core significantly smaller and less massive than ours. From the column density maps we have estimated the optical depth $\tau$ maps,  using the spectral parameters listed in the Cologne Database for Molecular Spectroscopy (CDMS\footnote{\url{https://cdms.astro.uni-koeln.de/}}), through the equation:
\begin{equation}
\label{tau}
\tau_\nu = \sqrt{\frac{\ln 2}{16 \pi^3}}   \frac{c^3 A_\mathrm{ul}  g_\mathrm{u} }{\nu^3 Q(\text{\tex}) FWHM} e^{- \frac{E_\mathrm{u}}{k_\mathrm{B} \text{\tex}}} \left(  e^{ \frac{h \nu}{k_\mathrm{B} \text{\tex}} } -1 \right) N_\mathrm{col} \; ,
\end{equation}
where $A_\mathrm{ul} = 1.20\times 10^{-8} \, \rm s^{-1}$ is the Einstein coefficient for spontaneous emission, $g_\mathrm{u} = 9$ is the statistical weight, $E_\mathrm{u} = 17.9 \, \rm K$ is the upper level energy, $Q(\text{\tex})$ is the partition function (which at $10\, \rm K$ is $Q = 10.48 $), $k_\mathrm{B}$ is the Boltzmann constant, and $c$ is the light speed. We report the average value of $\tau_\nu$ and its dispersion around the mean value in each core in Table \ref{CoreProp}. Most of the cores present $\tau \lesssim 1.0$, indicating that the \olineh line is optically thin. Less than 7\% of pixels in both clumps show optical depths higher than unity. This suggests that self-absorption is not affecting the line shapes strongly, even though we cannot completely exclude this hypothesis given the scarce spectral resolution of ALMA observations. 
\par
The velocity dispersion maps of the clumps show that ALMA detects very narrow \ohhdp lines. In several cores, especially in AG351, large portions of the gas traced by \hhdp present total linewidths narrower than the isothermal sonic speed, which at $T_\mathrm{gas} = 10 \, \rm K$ is\footnote{In the assumption of LTE continuum, the excitation temperature of the line equals the local gas kinetic temperature.}:
\begin{equation}
c_\mathrm{s} = \sqrt{\frac{k_\mathrm{B} T_\mathrm{gas}}{\mu m_\mathrm{H}}} =  0.19 \, \text{\kms} \; ,
\end{equation} 
where  $m_\mathrm{H}$ is the mass of the hydrogen atom, and $\mu$ is the mean molecular weight of the gas
($\mu = 2.33$ for a gas composition of H$_2$ and 10\% of helium). In particular, in AG351 36\% of positions show subsonic linewidths\footnote{Velocity dispersion values (total, thermal, non-thermal,...) are subsonic or supersonic depending on whether their ratio with respect to the sound speed is lower or higher than unity, respectively.}, whilst the percentage for AG354 is 23\%. The fraction of subsonic gas could be even higher, if we consider that the limited spectral resolution of our observations may cause an overestimation of the linewidths. This has profound implication on the physical conditions traced by \olineh, as discussed in detail in Sect. \ref{Disc:1}. Locally, the line velocity dispersion increases up to $0.4-0.5\,$\kms, without however showing a clear correlations with other quantities, such as the emission in continuum. \par
\begin{figure}[h]
\centering
\includegraphics[width=0.5\textwidth]{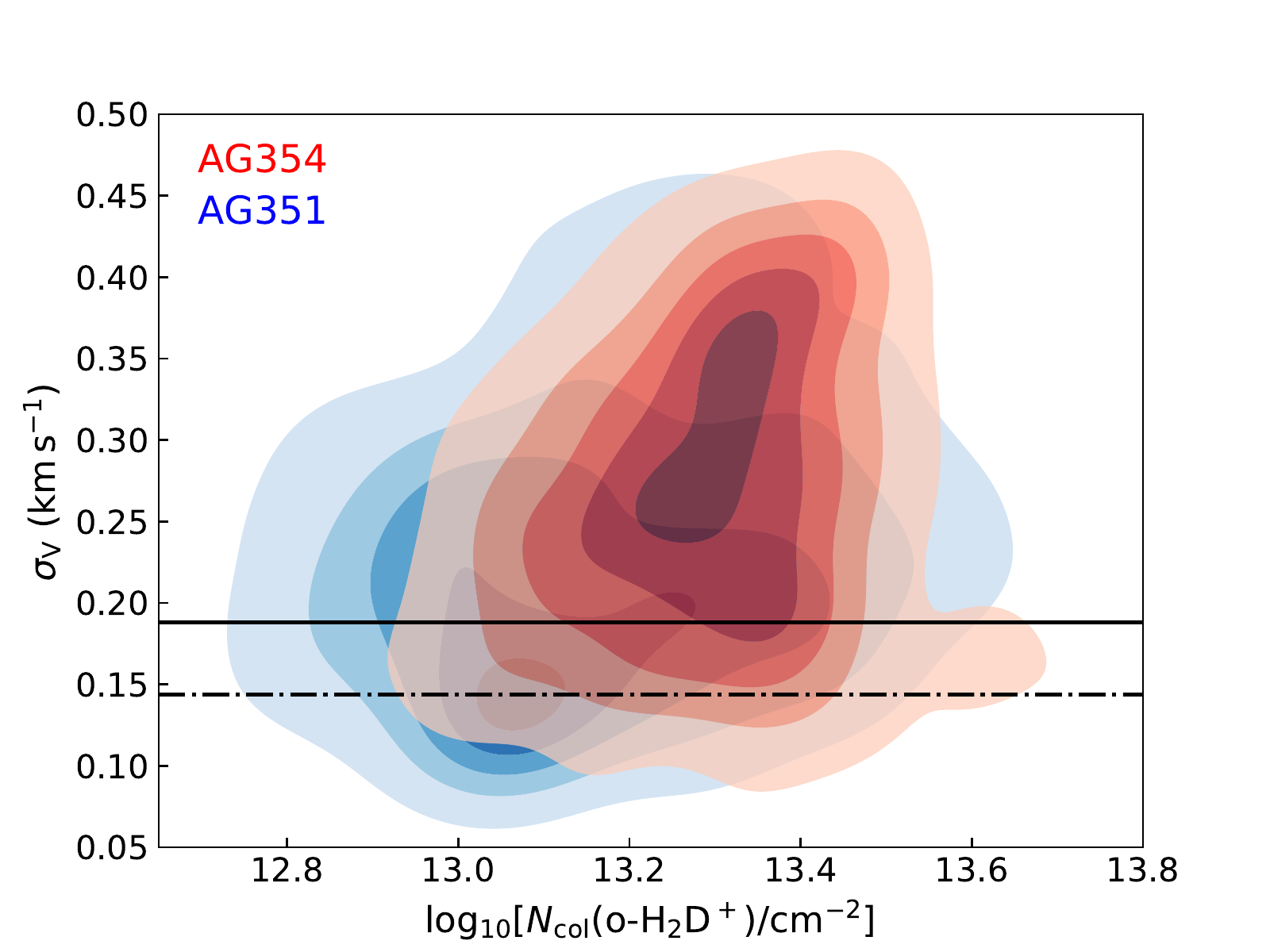}
\caption{Normalised kernel density distribution of \sigmav and \ncol in AG351 (blue) and AG354 (red) obtained from the \textsc{MCWeeds} pixel-per-pixel fit. The contour levels are $[0.1, 0.3, 0.5, 0.7, 0.9]$. The horizontal solid line shows the sound speed at $10\, \rm K$, whilst the dot-dashed line represents the thermal broadening of \hhdp at the same temperature. \label{Dens1} }
\end{figure}

We have compared the distribution of the velocity dispersion and column density in the two clumps in Fig. \ref{Dens1}. These plots show that AG351 presents in general narrower lines and lower column density values than AG354, which could hint to differences in their evolutionary stage and/or physical conditions.

\subsection{Average cores' parameters \label{AveParam}}
Important dynamical quantities of cores, such as the virial mass (see Sect. \ref{Disc:1} for further discussion), depend on their average properties. In order to derive these parameters, we have computed the average \olineh spectrum in each core. As highlighted in the previous subsection, the clumps present velocity gradients of the order of $1-2\,$\kms, while the individual cores show smaller variations of the centroid velocity. Nevertheless, given the narrow linewidths of the \ohhdp spectra, even these small gradients could affect the line shapes of the averaged spectra, if not taken into account. Before computing the mean spectrum from a core, therefore, we first aligned the spectra pixel-per-pixel to the mean centroid velocity obtained from the spectral fitting in Sect. \ref{MCWeeds}. \par
\begin{figure*}[h]
\centering
\includegraphics[width=0.9\textwidth]{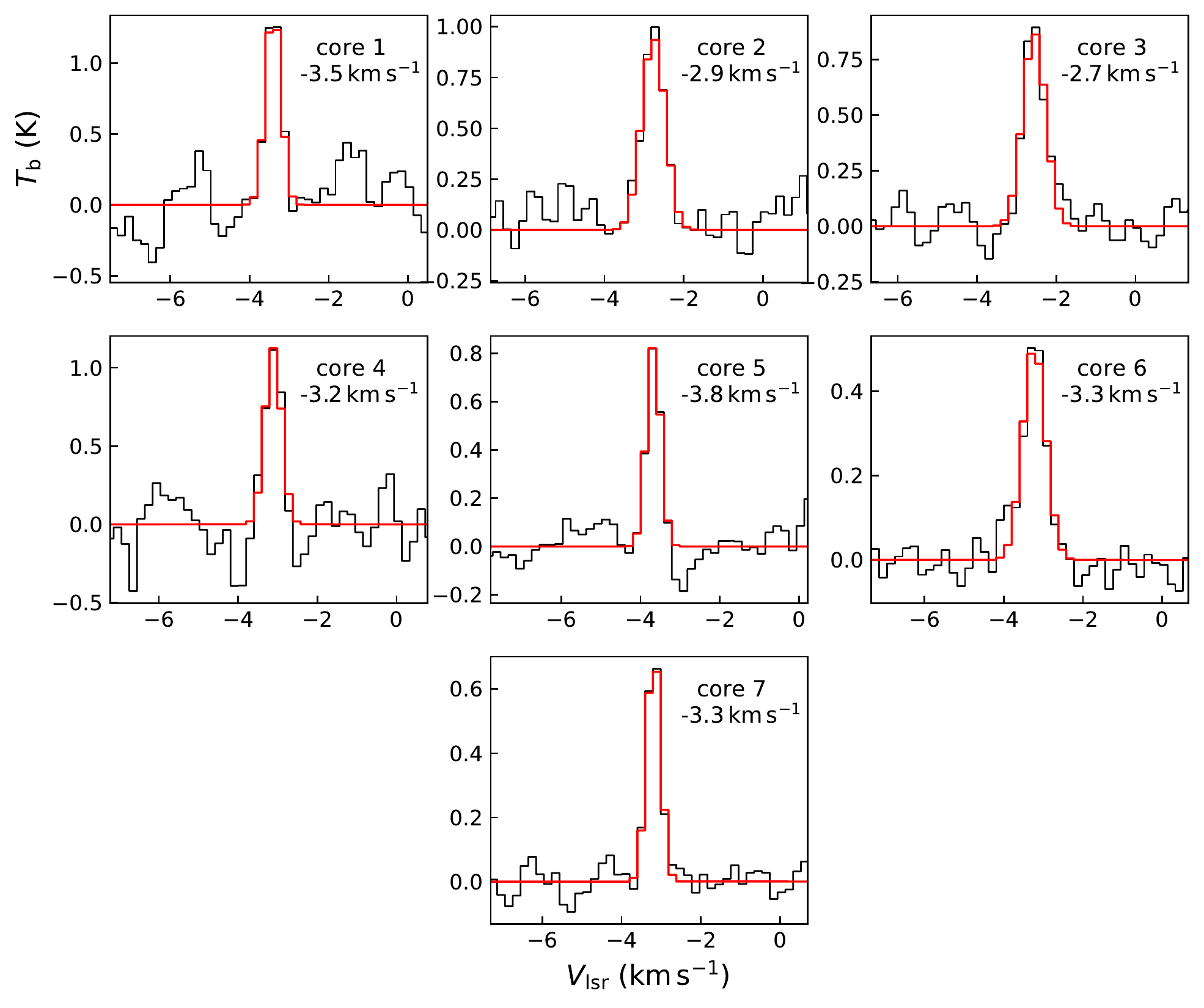}
\caption{The mean signal obtained averaging in each core the observed spectra is shown in black. The red curve shows the best-fit model obtained with \textsc{MCWeeds}. In the top-right corner of each panel we report the core id number and the \vlsr value to which the spectra have been aligned, before averaging them. \label{AG351_AveSp}}
\end{figure*}
\begin{figure*}[h]
\centering
\includegraphics[width=0.9\textwidth]{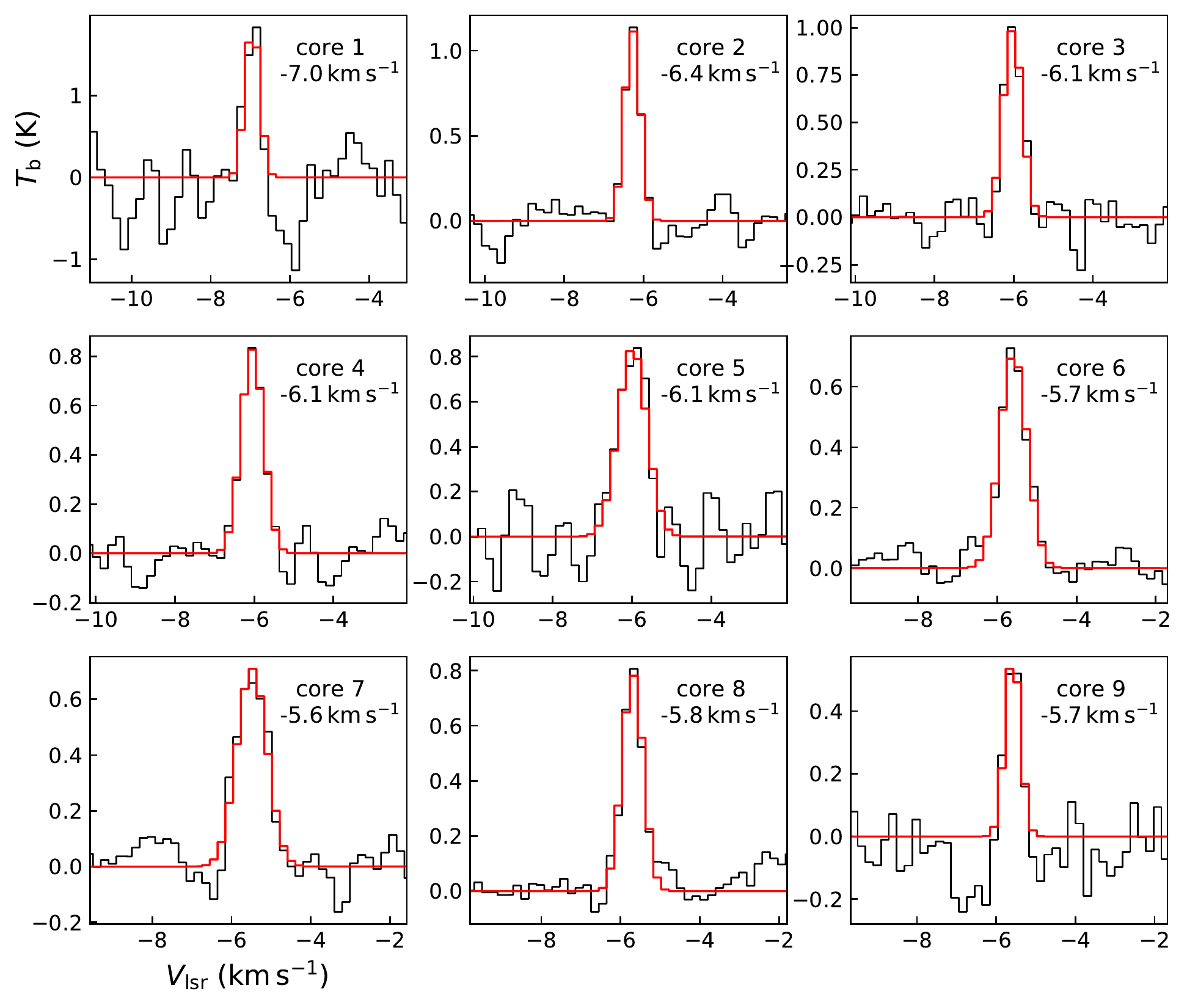}
\caption{Same as Fig. \ref{AG351_AveSp}, but for AG354. \label{AG354_AveSp} }
\end{figure*}

The averaged spectra have then been re-analysed using \textsc{MCWeeds}, as previously described, setting $\text{\tex} = 10\, \rm K$ and fitting the linewidth, centroid velocity, and column density of \ohhdp. We have computed the $rms$ values of the average spectra considering line-free channels. Table \ref{CoreProp} summarises the best fit values, together with their HPD intervals at 95\% level, obtained with the method just described.  We show the average spectra, overlaid with the best-fit model obtained with \textsc{MCWeeds}, in Figs. \ref{AG351_AveSp} and \ref{AG354_AveSp}. It is worth noting that all the spectra present Gaussian line shapes. This confirms first of all that secondary velocity components are in general negligible. Furthermore, it also suggests that opacity effects (such as self-absorption) are likely negligible as confirmed by the optical depth values computed in Sect. \ref{MCWeeds} (see Table \ref{CoreProp}), even though we cannot exclude them completely. In fact, due to the limited spectral resolution of the ALMA observations, the observed velocity dispersion may be locally overestimated, which in turns lead to the underestimation of the optical depth.\par
Dynamical parameters estimated from the observed velocity dispersions can provide useful information on the level of turbulence and its impact on the core dynamics. In particular, it is interesting to investigate whether the gas motions are sub- or super-sonic. The total velocity dispersion values derived from the average spectra (Table \ref{CoreProp}, 7th column) show that about half of cores present subsonic or trans-sonic gas motions, within uncertainties. We can further disentangle the thermal and non-thermal components of the lines. In fact, the velocity dispersion of any line is due to a combination of its thermal broadening and non-thermal one, the latter comprising all terms which are not due to the gas temperature, such as turbulence, multiple components on the line-of-sight if spectrally unresolved, and instrumental broadening. In the assumption that the two components are independent, they sum in quadrature to compose the total, observed velocity dispersion: $\sigma_\mathrm{V} = \sqrt{\sigma_\mathrm{V,NT} ^2 +\sigma_\mathrm{V, th} ^2 }$ \citep{Myers83}. The thermal component of the \ohhdp line at the gas temperature $T_\mathrm{gas}$ is:
\begin{equation}
\sigma_\mathrm{V,th} =  \sqrt{\frac{k_\mathrm{B} T_\mathrm{gas}}{m_\text{\hhdp} }} \; ,
\end{equation}
where $m_\text{\hhdp}$ is the \hhdp molecular mass in g. We have computed the ratio of non-themal velocity dispersion and sound speed for each core from the observed velocity dispersion, as obtained with the spectral fit, assuming $T_\mathrm{gas}= 10 \, \rm K$, at which $\sigma_\mathrm{V,th}=0.14\,$\kms. The one-dimension turbulent Mach number ($ \sigma_\mathrm{V,NT}/ c_\mathrm{s}$) values are summarised in Table \ref{CoreProp}. In AG351, all cores present $\sigma_\mathrm{V,NT}/ c_\mathrm{s}$ values lower than unity, or consistent with unity within 95\% HPD. Three cores in AG354, instead, present slightly supersonic turbulent motions within the confidence intervals. In a similar way, \cite{Sabatini20} found that at the clump scales the one-dimension turbulent Mach number in AG354 is higher than in AG351, even though according to their results both sources are mildly supersonic ($\sigma_\mathrm{V,NT}/ c_\mathrm{s} = 1.4$ and $1.8$ in AG351 and AG354, respectively). 

 \begin{table*}[h]
 \renewcommand{\arraystretch}{1.4}
\centering
\caption{Cores' properties and best-fit results obtained fitting their average spectra with \textsc{MCWeeds}. The $rms$ values are standard deviation over line-free channels. Uncertainties on \vlsr, \sigmav, \ncol, and one-dimension turbulent Mach number are expressed as 95\% high probability density intervals.}
\label{CoreProp}
\begin{tabular}{cccccccccc}
\hline
Core id      &  \multicolumn{2}{c}{Position}  & $R_\mathrm{eff}$      & $rms$       & \vlsr         &    \sigmav                       & \ncol     &  $\sigma_\mathrm{V,NT}/ c_\mathrm{s}$    & $\tau$\tablefootmark{a}     \\
	& 		RA  (h:m:s.ss)   &    Dec (d:m:s.ss) &	$10^{3}$AU	& mK 		& \kms & \kms				& $\log_{10} ( \rm cm^{-2})$	      &		&     \\
\hline \hline
\multicolumn{9}{c}{AG351}                                                                \\

1  	&17:20:50.23 & $-35$:35:24.26  	& 1.5 	& 160 & $-3.52^{+0.04}_{-0.05}$ & $0.18^{+0.05}_{-0.04}$ & $13.40^{+0.10}_{-0.08}$    & $ 0.6_{-0.4} ^{+0.4}$	&	$1.0 \pm 0.3 $ \\
2  	&17:20:51.75 & $-35$:35:23.15  	&1.4  	& 140 & $-2.88^{+0.05}_{-0.07}$ & $0.28^{+0.07}_{-0.07}$ & $13.39^{+0.09}_{-0.10}$  & $ 1.3_{-0.5} ^{+0.4}$	&	$ 0.63\pm 0.17  $ \\
3  	&17:20:51.57 & $-35$:35:25.35 	& 2.7   	& 100 & $-2.68^{+0.06}_{-0.05}$ & $0.27^{+0.07}_{-0.06}$ & $13.33^{+0.08}_{-0.07}$  & $ 1.2_{-0.4} ^{+0.4}$	&	$ 0.6\pm 0.2 $ \\
4  	&17:20:50.32 & $-35$:35:27.19  	&1.4   	& 180 & $-3.23^{+0.06}_{-0.08}$ & $0.19^{+0.06}_{-0.06}$ & $13.33^{+0.10}_{-0.13}$   & $ 0.7_{-0.6} ^{+0.4}$	&	$ 0.8\pm 0.2 $ \\
5   	&17:20:50.59 & $-35$:35:25.31 	& 2.7   	& 100 & $-3.79^{+0.02}_{-0.02}$ & $0.16^{+0.02}_{-0.02}$ & $13.11^{+0.04}_{-0.03}$   & $ 0.4_{-0.2} ^{+0.2}$	&	$ 0.49\pm 0.17 $ \\
6  	&17:20:50.98 & $-35$:35:25.17 	& 2.1    	& 40 & $-3.35^{+0.04}_{-0.04}$ & $0.28^{+0.05}_{-0.04}$ & $13.08^{+0.05}_{-0.05}$   & $ 1.3_{-0.3} ^{+0.3}$	&	$ 0.25\pm 0.06 $ \\
7	&17:20:51.31 & $-35$:35:23.67  	& 2.2   	& 40 & $-3.31^{+0.02}_{-0.02}$ & $0.17^{+0.03}_{-0.03}$ & $13.04^{+0.05}_{-0.04}$   & $ 0.4_{-0.3} ^{+0.2}$	&	$ 0.38 \pm 0.16 $ \\
\hline
\multicolumn{9}{c}{AG354}                                                                                                                        \\
1    	&17:35:11.64 & $-33$:30:18.37 	& 1.5  	& 400 & $-7.04^{+0.20}_{-0.10} $         & $0.16^{+0.10}_{-0.09}$         & $13.54^{+0.19}_{-0.23}$  & $ 0.5_{-0.4} ^{+0.6}$	&	$ 1.8\pm 0.5 $ \\
2  	&17:35:11.89 & $-33$:30:20.86 	& 2.9   	& 100   & $ -6.37^{+0.04}_{-0.04}$       & $ 0.18^{+0.04}_{-0.03}$        & $13.30^{+0.06}_{-0.07}$ & $ 0.6_{-0.4} ^{+0.3}$	&	$0.7 \pm 0.3 $ \\
3   	&17:35:11.64 & $-33$:30:35.62  	& 2.3    	& 130                     & $ -6.13^{+0.05}_{-0.06}$       & $ 0.20^{+0.06}_{-0.05}$        & $ 13.32^{+0.09}_{-0.10}$   & $ 0.8_{-0.5} ^{+0.4}$	&	$0.63  \pm 0.18 $ \\
4    	&17:35:12.15 & $-33$:30:34.63 	& 3.3   	& 70    & $ -6.15^{+0.04}_{-0.04}$       & $0.27^{+0.04}_{-0.04}$         & $13.30^{+0.06}_{-0.05}$   & $ 1.2_{-0.2} ^{+0.2}$	&	$0.49 \pm 150 $ \\
5  	&17:35:11.79 & $-33$:30:35.94  	& 1.5 	& 150     & $-6.09^{+0.09}_{-0.10}$        & $ 0.35^{+0.08}_{-0.07}$        & $13.42^{+0.10}_{-0.09}$  & $ 1.7_{-0.5} ^{+0.4}$	&	$0.6 \pm 0.2 $ \\
6	&17:35:11.96 & $-33$:30:30.79 	& 2.9  	& 50    & $-5.70^{+0.05}_{-0.03}$        & $0.33^{+0.05}_{-0.04}$         & $13.31^{+0.04}_{-0.04}$ & $ 1.6_{-0.2} ^{+0.3}$	&	$ 0.38\pm 0.07 $ \\
7    	&17:35:12.17 & $-33$:30:27.66  	& 1.9   	& 60  & $-5.59^{+0.04}_{-0.05}$        & $0.36^{+0.04}_{-0.04}$         & $13.36^{+0.05}_{-0.04}$  & $ 1.8_{-0.2} ^{+0.3}$	&	$ 0.39\pm 0.06 $ \\
8   	&17:35:12.24 & $-33$:30:30.24 	& 2.8 	& 40  & $-5.80^{+0.02}_{-0.02}$        & $0.23^{+0.03}_{-0.02}$         & $13.25^{+0.03}_{-0.03}$ & $ 0.95_{-0.16} ^{+0.16}$	&	$ 0.45 \pm 0.13 $ \\
9    	&17:35:12.28 & $-33$:30:24.65 	&  1.5  	& 80     & $-5.69^{+0.05}_{-0.05}$        & $0.18^{+0.05}_{-0.05}$         & $12.98^{+0.09}_{-0.11}$   & $ 0.6_{-0.5} ^{+0.4}$	&	$ 0.30 \pm 0.08 $ \\    
\hline                
\end{tabular}
\tablefoot{
\tablefoottext{a}{Optical depth of the \ohhdp transition, expressed as average value $\pm$ standard deviation around the average within each core. The values are computed assuming $T_\mathrm{ex}=10\, \rm K$.}}
\end{table*}

  \section{Discussion\label{Discussion}}
\subsection{The gas traced by \ohhdp\label{Disc:1}}
As explained in Sect. \ref{Introduction}, the \olineh transition probes a cold and dense component of the interstellar medium. However, having access to only one molecular transition prevents us to determine the gas temperature and density independently at the same time. In order to perform the spectral analysis, we have assumed an excitation temperature of \tex$=10\, \rm K$, which is consistent with previous works \citep{Harju06, Pillai12, Friesen14, Sabatini20} and with the fact that \hhdp is abundant when the temperature is $\lesssim 20\, \rm K$. However, dense prestellar gas can reach temperatures lower than $10\, \rm K$, as shown both by theoretical works \citep{Zucconi01, Evans01, Ivlev19} and by observational evidence \citep{Crapsi07, Pagani07}. \par
The linewidth of the \olineh transition can give us hints on the gas temperature. As mentioned in the previous Sect., the velocity dispersion of any line can be separated in thermal and non-thermal broadening, which are summed in quadrature. It is therefore straightforward that the total linewidth of a line cannot be smaller than its thermal component. The thermal broadening of \hhdp at $10\, \rm K$ is $\sigma_\mathrm{V,th} = 0.14 \, $\kms. From the maps of this quantity obtained in Sec. \ref{MCWeeds} it results that 17\% of pixels in AG351 and 7\% in AG354 show linewidths smaller than this level (see Fig. \ref{Dens1}). This is a clear indication that at least some parts of the gas traced by \ohhdp have a temperature lower that $10\, \rm K$. We can reverse this argument, and derive the gas temperature $T_\mathrm{H_2D^+}$ at which the observed velocity dispersion values are purely due to thermal broadening:
\begin{equation}
\sigma_\mathrm{V} =  \sigma_\mathrm{V,th} \quad \rightarrow \quad T_\mathrm{H_2D^+} = \left ( \sigma_\mathrm{V} \right)^2 \times 
 \frac{m_\text{\hhdp}}{k_\mathrm{B}} \; .
\label{Tth}
\end{equation}
We have hence derived maps of $T_\mathrm{H_2D^+}$ pixel-per-pixel in each source. This quantity represents an upper limit on the real gas temperature, since when any contribution from the non-thermal component is present it holds that $\sigma_\mathrm{V} > \sigma_\mathrm{V,th}$. This approach has already been used for instance by \cite{Harju08}, who estimated a gas temperature of $T_\mathrm{H_2D^+} = 6\, \rm K$ from APEX observations of \olineh in the starless clump Ophiuchus D. In our clumps, we constrain $T_\mathrm{H_2D^+}$ to be in the range $5-20\, \rm \,K$. The lower limit is due to the fact that \textit{i)} temperature lower than $5 \, \rm K$ are not predicted even at very high volume densities (see, for instance, \citealt{Hocuk17}) and \textit{ii)} this is the temperature at which the thermal linewidth (in $FWHM$) equals the spectral resolution of our observations. The upper limit is instead due to the fact that the \hhdp abundance is expected to drop at temperatures higher   {than} $20 \rm \,K$ (which corresponds to $\sigma_\mathrm{V,th}= 0.21 \,$\kms). Linewidths broader than this limit most likely involve non-thermal contributions, such as turbulence. A significant fraction of positions presents $T_\mathrm{H_2D^+} < 20\,\rm K$ (50\% in AG351 and 31\% in AG354).  \par
The quantities derived from the \textsc{MCWeeds} fit of the average spectrum in each core allow us to derive another important property: the virial mass ($M_\mathrm{vir}$), which can be used to assess the dynamical state of a source. This is the mass at which the kinetic energy content of a system equals its gravitational potential energy, and it can be expressed in observational units as \citep{Bertoldi92}:
\begin{equation}
M_\mathrm{vir} = \frac{5 R_\mathrm{core} \sigma_\mathrm{dyn}^2 }{G} =  1200 \times \left ( \frac{R_\mathrm{core}}{\mathrm{pc}} \right ) \left ( \frac{\sigma_\mathrm{dyn}}{\mathrm{km \, s^{-1}}} \right )^2 \rm{M_\odot} \; ,
\label{Mvir}
\end{equation}
where $G$ is the gravitational constant. Equation \eqref{Mvir} holds in the assumption of a spherical core of radius $R_\mathrm{core}$ and uniform density, and $\sigma_\mathrm{dyn}$ represents the total linewidth of the gas, and it can be derived from the observed \sigmav of \ohhdp assuming that the non-thermal component of the latter is the same for all the gas components:
\begin{equation}
\sigma_\mathrm{dyn} = \sqrt{\sigma_\mathrm{NT}^2 + c_\mathrm{s}^2 } \; .
\end{equation}
\par
In a first step, we estimate $M_\mathrm{vir}$ from the \ohhdp linewidths and the effective radii listed in Table \ref{CoreProp} ($R_\mathrm{core} = R_\mathrm{eff}$), using a constant gas temperature of $10\, \rm K$. For each core we compute the distribution of the virial mass from the corresponding distribution of \sigmav, and we then determine the median and the 95\% HPD intervals. The resulting virial masses are summarised in Table \ref{CoreDyn}. \par
As previously discussed, we have reasons to believe that at least part of the gas traced by \ohhdp is colder than $10\, \rm \, K$. As a test, we have therefore computed the virial masses using the minimum $T^\mathrm{min}_\mathrm{H_2D^+}$ value in each core as derived from Eq. \eqref{Tth}. With some little algebra it can be shown that $\sigma_\mathrm{dyn}$  (and hence $M_\mathrm{vir}$) is positively correlated with temperature. Since  $T_\mathrm{H_2D^+}$ represents an upper limit for the gas temperature, the virial masses derived in this way also represent upper limits. In Table \ref{CoreDyn} we also report the $M_\mathrm{vir}$ values obtained with this approach, together with the corresponding  $T_\mathrm{H_2D^+}$ values used for the computation. The virial mass does not strongly depend on the gas temperature, and the new $M_\mathrm{vir} $ values differ by only a few \% from the previous ones. It is interesting to notice that the estimated virial masses are in the range $0.3-2.2\, \rm M_\odot$, and hence these cores are essentially low-mass, in the hypothesis that they are virialised by the kinetic energy. We further discuss this point in Sec. \ref{Disc:2} and \ref{Disc:4}.

 \begin{table*}[h]
 \renewcommand{\arraystretch}{1.4}
\centering
\caption{Properties of the identified cores: virial mass, total mass from dust continuum emission, and average volume density. The first set of values is computed in the assumption that $T_\mathrm{gas} =T_\mathrm{dust} =10\, \rm K$; in the second half of the table, the quantities are computed using the minimum temperature derived from the \hhdp thermal broadening. The used temperature values are listed in the 5th column.}
\label{CoreDyn}
\begin{tabular}{c|cccc|ccccc}
\hline
 		&   \multicolumn{4}{c|}{ $T= 10\,\mathrm{K = const}$}  &   \multicolumn{5}{c}{$T =  T^\mathrm{min}_\mathrm{H_2D^+}$}  \\ 
Core id  	& 	$M_\mathrm{vir} $   & $M_\mathrm{core}$\tablefootmark{a} & $n \rm (H_2)$   & $\alpha_\mathrm{vir}$ &  $ T^\mathrm{min}_\mathrm{H_2D^+}$\tablefootmark{b}&$M _\mathrm{vir} $   & $M _\mathrm{core}$\tablefootmark{a} & $n\rm  (H_2)$ & $\alpha _\mathrm{vir}$\\ 
		& 	$\mathrm{M_\odot}$	&	$\mathrm{M_\odot}$	&	$10^6 \rm cm^{-3}$	&		& K&	$\mathrm{M_\odot}$	&	$\mathrm{M_\odot}$	&	$10^6 \rm cm^{-3}$		&			\\
\hline \hline
\multicolumn{10}{c}{AG351}                                                                \\
 1&	$ 0.39_{-0.12} ^{+0.16}$	&$0.7\pm 0.2 $ &  6.2  &$0.58_{-0.18}^{+0.23}$ 		&5.0	&	$0.32_{-0.12} ^{+0.16}$	&	$4.7\pm 1.5 $	& 42.8	& $0.07_{-0.03}^{+0.03}$	 \\ 
 2&	$ 0.7_{-0.3} ^{+0.4}$		&$0.53\pm 0.17 $ &  6.4  & $1.4_{-0.5}^{+0.7}$ 			&6.1	&	$0.7_{-0.0.3} ^{+0.04}$	&	$1.9\pm 0.6$	& 22.9	& $0.36_{-0.14}^{+0.19}$ 	 \\ 
 3&	$ 1.0_{-0.3} ^{+0.5}$		&$1.8\pm 0.6$ &  7.6  & $0.54_{-0.18}^{+0.26}$ 		&5.0	&	$0.9 _{-0.3} ^{+0.5}$		&	$13\pm 4$	& 52.6	& $0.07_{-0.03}^{+0.04}$ 	 \\ 
 4 &	$ 0.4_{-0.2} ^{+0.2}$		&$0.43\pm 0.14 $ &  5.1  &	$0.9_{-0.4}^{+0.5}$ 		&5.4	&	$0.35_{-0.16} ^{+0.20}$	&	$2.2\pm 0.7 $	& 26.3	&$0.16_{-0.08}^{+0.09}$ 	\\ 
 5 &	$0.63_{-0.08} ^{+0.09}$	&$1.6\pm 0.5 $ &  2.3  & $0.40_{-0.05}^{+0.06}$ 		&5.0	&	$0.52_{-0.08} ^{+0.09}$	&	$11\pm 4 $	& 16.1 	& $0.05_{-0.01}^{+0.01}$	 \\ 
 6 &	$ 1.1_{-0.3} ^{+0.4}$		&$1.1\pm 0.4 $ &  3.5  & $1.0_{-0.2}^{+0.3}$ 			&11.0&	$1.1_{-0.3} ^{+0.4}$		&	$0.9\pm 0.3 $	& 2.9		& $1.2_{-0.3}^{+0.4}$ 		 \\ 
 7 &	 $ 0.51_{-0.10} ^{+0.10}$	&$1.6\pm 0.5 $ &  4.6  & $0.32_{-0.06}^{+0.06}$ 		&5.0 	&	$0.42_{-0.10} ^{+0.09}$	&	$11\pm 4 $	& 32.2	& $0.04_{-0.01}^{+0.01}$ 	 \\ 
\hline
\multicolumn{10}{c}{AG354}                 \\
 1   &$ 0.35_{-0.18} ^{+031}$	&	- 	            &	-      & - 					&5.0	&	$0.28_{-0.18} ^{+0.31}$	&		-  		        & - 		& -					\\
 2   &$ 0.8_{-0.2} ^{+0.2}$ 	&$5.6\pm 1.8 $ &  7.3  & $0.14_{-0.04}^{+0.04}$	&5.0	&	$0.65_{-0.20} ^{+0.23}$	&   	$39\pm 13$	& 50.7	& $0.02_{-0.01}^{+0.01}$ 	\\ 
  3 &$ 0.7_{-0.3} ^{+0.3}$		&$1.1\pm 0.4 $ &  2.8  & $0.6_{-0.2}^{+0.3}$		&6.2	&	$0.7_{-0.3} ^{+0.3}$		&	$3.8\pm 1.3 $	& 9.4	 	& $0.17_{-0.07}^{+0.09}$	\\ 
 4   &$ 1.6_{-0.4} ^{+0.5}$		&$3.0\pm 1.0 $ &  2.5  & $0.54_{-0.12}^{+0.15}$	&8.5	&	$1.6_{-0.4} ^{+0.5}$		&	$4.3\pm 1.4 $	& 3.9		& $0.36_{-0.08}^{+0.11}$ 	\\ 
 5 &$ 1.1_{-0.4} ^{+0.5}$		&$0.38\pm 0.12 $ &  3.6  &$3.0_{-1.0}^{+1.3}$ 		&20.0&	$1.2_{-0.4} ^{+0.5}$		&	$0.11\pm 0.04 $	& 1.0		& $11.2_{-3.3}^{+4.4}$  	\\ 
 6 &	$2.0_{-0.4} ^{+0.5}$		&$2.2\pm 0.7 $ &  2.7  & $0.90_{-0.17}^{+0.24}$	&20.0&	$ 2.2_{-0.4} ^{+0.5}$		&	$0.7\pm 0.2$	& 0.8		& $3.4_{-0.6}^{+0.8}$ 	\\ 
 7 &$ 1.5_{-0.3} ^{+0.4}$		&$1.17\pm 0.17 $ &  5.5  & $1.3_{-0.3}^{+0.3}$		&20.0&	$1.7_{-0.3} ^{+0.4}$		&	$0.34\pm 0.05 $	& 1.6	 	& $5.0_{-1.0}^{+1.1}$	\\ 
 8 &$ 1.10_{-0.16} ^{+0.18}$	&$2.1\pm 0.7 $ &  3.0  & $0.50_{-0.08}^{+0.09}$	&7.9	&	$1.02_{-0.16} ^{+0.18}$	&	$3.7\pm 1.2 $	& 5.1		& $0.28_{-0.04}^{+0.05}$ \\ 
 9 &$0.40_{-0.12} ^{+0.18}$	&	-		    &	-      & - 					&6.8	&	$0.36_{-0.12} ^{+0.18}$	&		-                        & - 		& -					\\ 
\hline                
\end{tabular}
\tablefoot{ 
\tablefoottext{a}{The uncertainties on the masses estimated from the continuum emission are 33\% (relative error), to take into account uncertainties on parameters such as the dust opacity, on top of the flux error from the ALMA observations.} \\
\tablefoottext{b}{$ T_\mathrm{H_2D^+}$ is the minimum $T_\mathrm{H_2D^+}$ in each core computed through Eq. \ref{Tth}. It still represents, however, an upper limit on the real gas/dust temperature, since other components (non-thermal, instrumental,...) can contribute to the line broadening.} }

\end{table*}

\subsection{The correlation between molecular and continuum emission\label{Disc:2}}
Our ALMA data allow us to compare the gas properties as traced by the dust thermal emission and by the \ohhdp transition. As previously mentioned, the two sets of data do not present the same morphology. We investigate this point in detail in Appendix \ref{App:contCores}, where we identify core-like structures in the continuum maps using dendrogram analysis. Out of the 16 \hhdp-cores, five do not overlap at all with continuum-identified structures, and conversely three structures identified in continuum emission do not find correspondence in the \textsc{scimes} analysis. Furthermore, for two \hhdp-identified cores the continuum flux peak is found outside the core boundaries. It is hence clear that the two datasets are not tracing in general the same material in the clumps.
In order to further investigate this point, we have used the continuum maps to estimate the total gas mass and density. In conditions of high densities ($n \gtrsim 10^{4-5}\rm \, cm^{-3}$), dust and gas are well coupled, which means that they are thermalised at the same temperature \citep{Goldsmith01}. We can hence assume $T_\mathrm{dust} = T_\mathrm{gas} = 10 \, \rm K$ in order to compute the total gas column density $N(\rm H_2)$ and hence the mass of the identified cores. We compute pixel-by-pixel the quantity:
\begin{equation}
N\mathrm{(H_2)} = f \frac{D^2 S_\mathrm{pix}}{B_\nu(T_\mathrm{dust}) \kappa_\nu A_\mathrm{pix} \mu_\mathrm{H_2} m_\mathrm{H} } \; ,
\label{NH2}
\end{equation}
where $f$ is the gas-to-dust ratio (assumed to be 100, \citealt{Hildebrand83}); $D$ is the source's distance; $B_\nu(T_\mathrm{dust})$ is the Planck function at the frequency $\nu$ and temperature $T_\mathrm{dust}$; $S_\mathrm{pix}$ is the pixel flux (in $\rm Jy \, pix^{-1}$) and $ A_\mathrm{pix}$ is the pixel area; $\mu_\mathrm{H_2} = 2.8$ is the mean molecular
weight per hydrogen molecule \citep{Kauffmann08}, and $ \kappa_\nu$ is the dust opacity at the frequency of the observations. For the latter, we use the power-law expression:
\begin{equation}
\kappa_\nu = \kappa_0 \left( \frac{\nu}{\nu_0}\right)^\beta = 1.71 \, \rm  cm^2 \, g^{-1} \; , 
\end{equation}
in which we employ $\beta = 1.5$ for the dust emissivity index \citep{Mezger90,Walker90} and the $\kappa_0 =10\, \rm  cm^2 \, g^{-1}$ for the dust opacity at the reference frequency corresponding to the wavelength $\lambda_0 = 250 \, \mu \rm m$ \citep{Hildebrand83,Beckwith90}. Errors on $N(\rm H_2)$ are computed with standard propagation from the uncertainties on continuum flux, and the average uncertainty is $rms=2.9 \times 10^{22} \, \rm cm^{-2}$. We have computed the average $N(\rm H_2)$ values for each core, which span the range $(1-4) \times 10^{23} \, \rm cm^{-2}$. From the gas column density we derive the molecular abundance $X_\mathrm{mol} (\text{\ohhdp}) = N_\mathrm{mol}(\text{\ohhdp})/ N(\mathrm{H_2})$.   {The derived values} are found in the range $(0.2-3)\times 10^{-10}$ for AG351 and $(0.1-10)\times 10^{-10}$ for AG354. These values are consistent with previous measurements. For instance, in low-mass prestellar and protostellar cores, \cite{Caselli08} found $\text{\xmol} =( 0.1 - 5)\times 10^{-10}$.  \par
\begin{figure*}[h]
\centering
\includegraphics[width=0.9\textwidth]{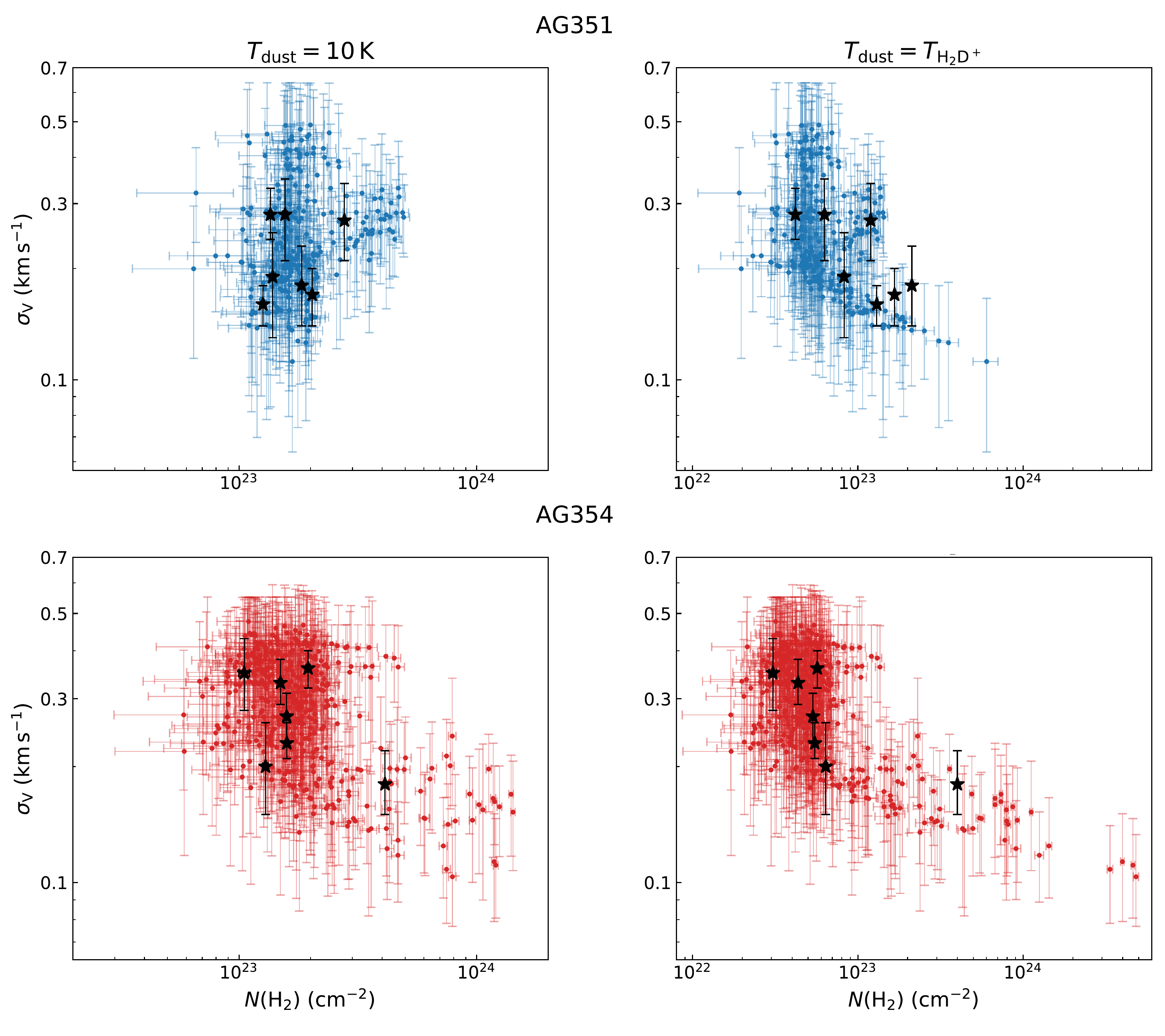}
\caption{Scatterplots of \olineh velocity dispersion $\sigma_\mathrm{V}$ with respect to $\rm H_2$ column density $N \rm (H_2)$ in AG351 (blue, top panels) and AG354 (red, bottom panels). In the left panels, $N\rm (H_2)$ is computed assuming a constant temperature of 10$\,$K. In the right panels, we use the temperature derived pixel-per-pixel from the \olineh linewidth, in the hypothesis of only thermal contributions. In order to allow an easier comparison between the two clumps, the plot ranges are kept equal in the corresponding panels. For the sake of readability, we only show data-points for which the relative 95\% HPD interval derived with \textsc{MCWeeds} is $<50$\%. The black stars show the average values referring to the \hhdp-identified cores, for which the continuum flux is significantly detected. The \sigmav values are taken from Table \ref{CoreProp}, whilst the $N\rm (H_2)$ values are computed as averages in each core. \label{Scatt1} }
\end{figure*}
\begin{figure*}[h]
\centering
\includegraphics[width=0.9\textwidth]{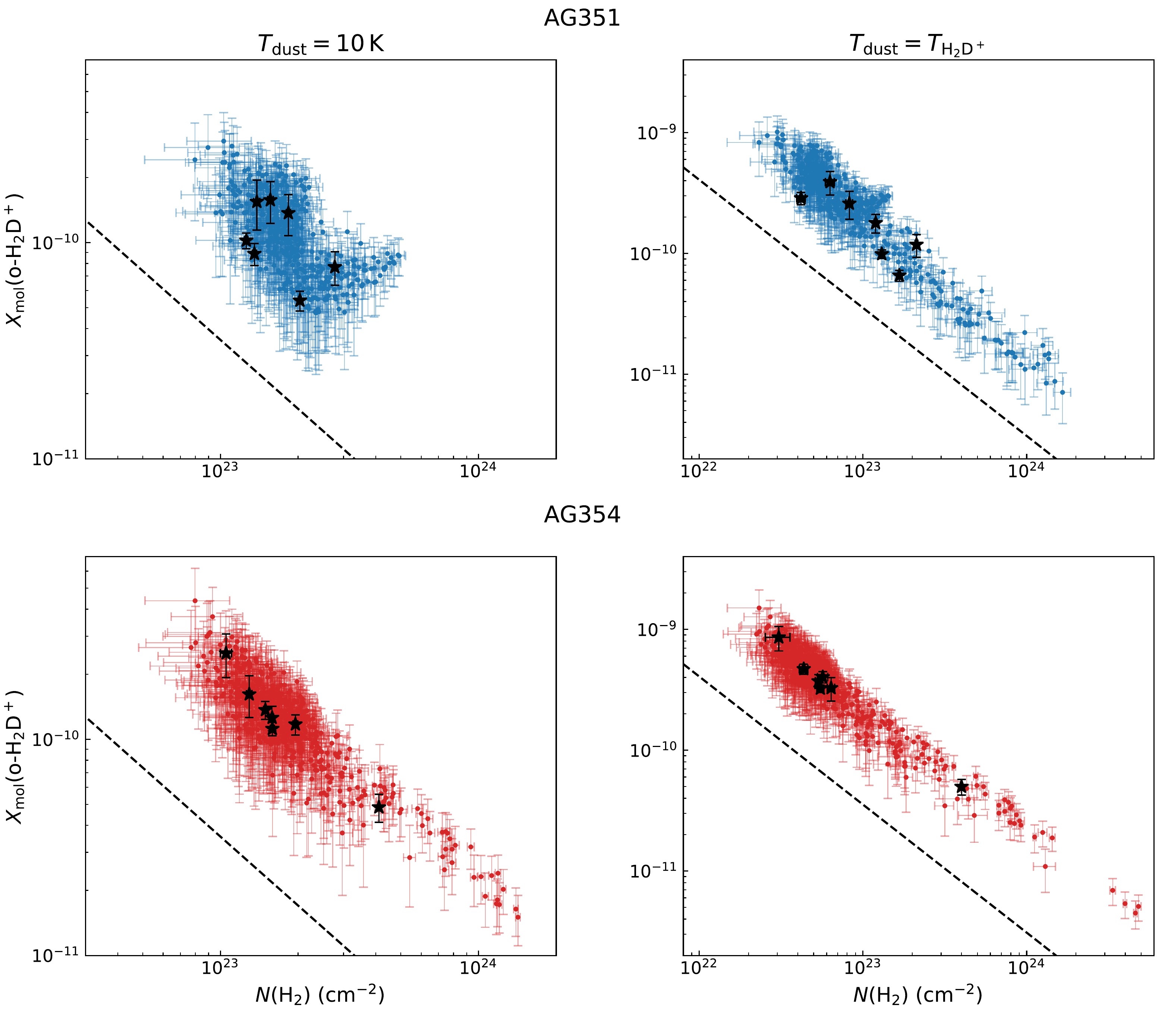}
\caption{Scatterplots of \ohhdp molecular abundance  with respect to $\rm H_2$ column density in AG351 (blue, top panels) and AG354 (red, bottom panels). In the left panels, the quantities are computed assuming a constant temperature of 10$\,$K. In the right panels, we use pixel-per-pixel the temperature derived from the \olineh linewidth, in the hypothesis of only thermal contributions. Scatterplot ranges are fixed to allow an easier comparison. For the sake of readability, we only show data-points for which the relative 95\% confidence interval is $<50$\%. The dashed, black curve shows the correlation found by \cite{Sabatini20}. The black stars show the average values referring to the \hhdp-identified cores, for which the continuum flux is significantly detected. The \ncol values to compute \xmol are taken from Table \ref{CoreProp}, whilst the $N\rm (H_2)$ values are computed as averages in each core. \label{Scatt2} }
\end{figure*}

 The scatterplots of the \ohhdp linewidth and molecular abundance with respect to $\mathrm{H_2}$ are shown in Figs. \ref{Scatt1} and \ref{Scatt2} (left panels), where we also show the average values for each core with star symbols. Concerning the relation between \sigmav and the $\rm H_2$ column density, no clear correlation is visible for AG351, while a tentative trend can be seen in AG354, where it seems that positions at high column densities ($N \rm (H_2) \gtrsim 5 \times 10^{23} \, cm^{-2}$) are characterised by narrow linewidths ($\sigma_\mathrm{V}\lesssim 0.35 \,$\kms).  The molecular abundance seems anti-correlated with respect to $N(\mathrm{H_2})$ for AG354, whilst the data do not show a clear trend in AG351. These differences in the scatterplots for the two clumps suggest that they are in two distinct evolutionary stages, as possibly suggested also by the fact that AG354 shows on average larger velocity dispersion and \ohhdp column density values. However, the long tail towards high gas densities in AG354 arises from the data points of one core mainly (core 2), and hence we cannot derive strong conclusions regarding the whole clump based on this.  
 \par
 The anticorrelation between $X_\mathrm{mol}$ and $N(\rm H_2)$ can be explained by the fact that at later evolutionary stages several factors can contribute to lowering the \hhdp abundance: depletion of HD onto the dust grains \citep{Sipila13}, conversion of \hhdp into its doubly and triply deuterated isotopologues, or a rise in temperature due to the protostellar feedback. Regarding the latter point, we lack the interferometric data to verify whether any protostellar source is embedded in the clumps. However, the fact that the observed velocity dispersion values in AG354 are higher than those of AG351 could be consistently explained by higher gas temperatures.  \par
From the continuum emission one can derive the total gas mass of the cores identified in \hhdp, using:
\begin{equation}
M_\mathrm{core} = f \frac{D^2 S_\mathrm{tot}}{B_\nu(T_\mathrm{dust}) \kappa_\nu } \; ,
\label{Mdust}
\end{equation}
where the notation is the same of Eq. \eqref{NH2}, with the exception that $S_\mathrm{tot}$ is now the total flux (in Jy) integrated over the area of the core. The obtained values are summarised in Table \ref{CoreDyn}. We have computed \mcore for cores where the peak flux is detected above the $5\sigma$ level. The ALMA absolute flux uncertainty in Band 7 is $\approx 10$\%. However, several parameters involved in the computation of \mcore are not well constrained, as for instance the dust opacity. A variation of 30\% of $\kappa_\nu$ translates in a variation of 33\% in the mass. We have therefore increased the error on \mcore to this level, in order to take into account the parameters uncertainties. \par 
The obtained masses range between 0.4-6 M$_\odot$, with larger values observed in AG354. However, these values might not represent the entire mass budget of the \hhdp-identified cores, particularly for the ones where the continuum emission is weak and anti-correlated with the molecular one. Furthermore, the interferometer may filter out the larger scale envelope emission, hence underestimating the cores' total masses. \par

Similarly to what discussed for the gas temperature, the assumption of a constant dust temperature $T_\mathrm{dust} = 10 \, \rm K$ is questionable. We do not have access to other data to better constrain this quantity. However, based on the assumption of dust-gas coupling, it holds that if the \ohhdp emission traces part of the gas colder than $10\, \rm K$, locally also the dust temperature can be lower than this value. We have therefore re-computed the $N\rm (H_2)$ (and hence the \ohhdp abundance) pixel-per-pixel using the gas temperature maps $T_\mathrm{H_2D^+}$ obtained from the \olineh linewidths, as discussed in Sec. \ref{Disc:1}. The average uncertainties on the new $N \rm (H_2)$ values are $rms=2.1 \times 10^{22} \, \rm cm^{-2}$ for AG351 and $rms=1.4 \times 10^{22} \, \rm cm^{-2}$ for AG354, and they are computed as standard error propagation from the flux uncertainty. The core average column density values are $N\mathrm{(H_2)} = 0.3-4 \times 10^{23} \, \rm cm^{-2}$. The new scatterplots showing the correlation of \sigmav and $X _\mathrm{mol} (\text{\hhdp})$ with respect to $N \rm  (H_2)$ are shown in the right panels of Figs. \ref{Scatt1} and \ref{Scatt2}. With the updated values, a trend between the observed velocity dispersion values and the gas total column density is visible for both sources. This is however expected, since the narrower the line, the lower the derived $T_\mathrm{H_2D^+}$, and hence the higher $N (H_2)$ (since, for a given flux, dust temperature and mass/density are anti-correlated). On the other hand, the anti-correlation between the molecular abundance and the $\rm H_2$ column density is now clear in both clumps. \par
The anti-correlation between continuum and deuterated species has already been observed in the literature. For instance, \cite{Zhang20} found a shift between $\rm NH_2D$ and dust thermal emission in several high-mass star forming regions. Concerning \ohhdp, \cite{Friesen14}  similarly reported a shift between the molecular emission peak and the continuum peak. \cite{Sabatini20} used the detections of \olineh in a sample of ATLASGAL sources to fit the relation between \xmol and $N \rm (H_2)$, which in the log-log scale reads: $\log_{10} [X\mathrm{_{mol}(\text{\ohhdp}})] = -1.06 \times \log_{10} [N\mathrm{(H_2)}] + 13.93$. We plot the predictions from this relation in Fig. \ref{Scatt2}. We have fit a linear relation in the log-log space between the molecular abundance and the $\rm H_2$ column density for our data. We find slopes of $m \approx -0.8,  -0.9$ with high correlation coefficient ($\left | R \right |  > 0.85$), with the exception of the \xmol-$N \rm (H_2)$ relation in AG351 with $T_\mathrm{dust} = 10\, \rm K$, where the data do not present a significant trend. The relations for our datasets show similar slopes to that found by \cite{Sabatini20}, but they are shifted upwards by a factor of $5-10$. Interestingly, the cores identified by \textsc{scimes} are on average $2.5''$ wide in angular size. Hence, the smaller abundances of \cite{Sabatini20} can be explained by beam dilution, if one takes into consideration the ratio of the source sizes with respect to the APEX beam, assuming $\approx 10$ cores in each clump: $16.8''^2/(2.5''^2 \times 10) \approx 5$. \par
It is important to notice that \xmol and $N \rm (H_2)$ are not independent variables. This point has been investigated deeply by \cite{Sabatini20}, who performed a robust statistical analysis of the correlation between these two quantities. In this work, we are more interested in the general trend that \xmol shows in relation to the gas total density, and in comparing our results with the literature ones. However, to further test whether the correlations that we see in Fig. \ref{Scatt2} are real, we have checked the correlation between the molecular column density $\text{\ncol}(\text{\ohhdp})$ and $N \rm (H_2)$ (not shown here). These two quantities are in fact independent\footnote{  We highlight that this is not entirely true, since both quantities are computed assuming a temperature value, and hence they both depend on this third variable. They should be analysed via a complete partial-correlation test, which however is beyond the scopes of this work.}. We do not find any significant correlation between these variables, as previously noted by \cite{Sabatini20}. The \ohhdp column density is flat with respect to $N \rm (H_2)$. This confirms that higher $N \rm (H_2)$ values present lower molecular abundances. 
\par

We have computed the core masses using the  $T\rm ^{min}_\mathrm{H_2D^+}$ of each core, as we did for the virial masses. We stress again that this temperature value is the minimum  $T_\mathrm{H_2D^+}$ value found within each core, but it represents an upper limit, due both to possible non-thermal components to the linewidth, and to the instrumental broadening caused by the limited spectral resolution of the observations. Hence, in reality the cores masses could be higher than these new $M\rm _{core} $ values, which are listed in Table \ref{CoreDyn}. It is interesting to notice that according to the dust masses computed at $10\, \rm K$, the cores are essentially low-mass. Only in the assumption of very low dust temperatures ($T_\mathrm{dust} \approx 5\, \rm K$), the estimated masses become larger, but in any case $M_\mathrm{core} \lesssim 13 \rm \, M_\odot$, with the only exception of core 2 in AG354. This holds also for the cores identified in the continuum emission maps in Appendix \ref{App:contCores}. Out of the 15 continuum-identified structures, only two are more massive than $20\, \rm M_\odot$ in the assumption of $T_\mathrm{dust} = 5\, \rm K$, one of which however do not present significant \ohhdp emission, hence preventing us from classifying it as truly prestellar. We further discuss these findings in Sec. \ref{Disc:4}. It is worth noting though
the lack of total-power observations means that the ALMA data might be filtering out the large scale emission associated with the gas surrounding the cores here identified, hence leading to underestimation of their masses based on the dust thermal emission (see Appendix \ref{APEX} for more details). 
 \par
From the dust masses and effective radii we can estimate the average volume density of the cores, in the assumption of constant distribution and spherical geometry, using the equation: 
\begin{equation}
 {n\mathrm{(H_2)} =  \frac{ 3 M_\mathrm{core}}{ 4 \pi R_\mathrm{eff}^3 \mu_\mathrm{H_2} m_\mathrm{H} } \; .}
\end{equation}
We have evaluated $n\rm (H_2)$ for both cases: assuming $T_\mathrm{dust} = 10 \, \rm K$ and using $T_\mathrm{dust} =   T\rm ^{min}_\mathrm{H_2D^+}$ derived from the \olineh linewidth. The resulting values are also summarised in Table \ref{CoreDyn}. Given the uncertainties of this estimation (e.g. due to the assumption of uniform, spherical distribution), we expect these values to be accurate within a factor of 2. In all cases, even when the dust mass is computed at a high temperature of $20\, \rm K$, the average density is higher than $10^6 \, \rm cm^{-3}$. This confirms that the ALMA data are tracing high-density material, and justifies the hypothesis of dust-and-gas coupling. Furthermore, at such high densities the \olineh is expected to be thermalised, and hence LTE is a good approximation \citep{Harju08}. The found volume densities correspond to surface densities in the range $\Sigma = (0.1-10.0 )\, \rm g \, cm^{-2}$, depending on the core and on the assumed temperature. \par
The ratio between the virial mass and the total mass, known as virial parameter ($\alpha_\mathrm{vir} = M_\mathrm{vir}/ M_\mathrm{core}$), provides indication on the dynamical state. For sub-virial sources ($\alpha_\mathrm{vir}< 1$), the kinetic energy content is not enough alone to balance the gravitational pull, and if no other source of pressure is present, they will collapse. On the other hand, supervirial sources ($\alpha_\mathrm{vir}> 1$) are unlikely to undergo gravitational contraction. We have computed the virial ratio in both cases, with a constant temperature of $10\, \rm K$ and with $T =  T^\mathrm{min}_\mathrm{H_2D^+}$, and we report the two sets of values in Table \ref{CoreDyn}. Most cores are subvirial, or consistent with being subvirial within the 95\% HPD intervals in AG351, regardless of the assumed temperature. In AG354, the supervirial cases are represented by core 5, 6, and 7 (i.e. 33\% of the sample). This is consistent with what found by \cite{Sabatini20} at the clump scales, who reported that AG351 is more subvirial than AG354 ($\alpha_\mathrm{vir} = 0.4$ and $ 0.8$, respectively). We cannot exclude however than the cores are virialised by sources of pressure other than the kinetic one, and in particular by the presence of magnetic fields. This point is further discussed in Sect. \ref{Disc:4}.

\subsection{A population of cores at different evolutionary stages\label{Disc:3}}
We can speculate that the different correlation between the dust continuum emission and the \hhdp distribution in the cores is linked to their evolutionary stage. In fact, \hhdp forms when the gas become dense and cold enough for reaction \eqref{Deuteration} to be efficient. As the gas becomes denser due to contraction motions, several factors could contribute to lowering the \hhdp abundance.  \hhdp can be transformed in $\rm D_2H^+$ and $\rm D_3^+$ or deplete as a consequence of HD depletion \citep{Sipila13}. Secondly, one has to take into account opacity effects: if the volume density grows much larger than the line critical density, the transition becomes optically thick ($\tau > 1$), which means that it does not trace anymore the whole bulk of gas, but only the layer at $\tau \approx 1$. In Sect. \ref{MCWeeds} we have derived the optical depths from the derived \ncol maps, and despite being a minority, there are positions where lines are moderately optically thick. This could contribute to explain why the observed velocity dispersion values in AG354, which we believe to be more evolved, are broader than in AG351. In fact, optically-thick lines deviate from gaussian shapes, which leads to the overestimation of their widths. When the prestellar core collapses and forms a protostellar object, the abundance of \hhdp is finally reduced by the rising temperature. \par
We can therefore speculate that cores with strong \hhdp emission, either corresponding to a continuum-identified structure or not, are early in their evolution and truly prestellar. The lack of detected continuum peaks and associated continuum-identified cores could be due to observational limits. In fact, if the dust temperature in the correspondence of the \ohhdp emission is lower than in the surroundings ---as expected in case of prestellar cores--- we may not be able to detect a centrally concentrated structure at $0.8\, \rm mm$ continuum emission, as shown for instance by \cite{DiFrancesco04}. Hence cores which do not have a correspondence in continuum-identified structures could still be centrally peaked in reality. \par
As the cores evolve, the \hhdp abundance at the dust continuum peak is lowered for the aforementioned reasons (e.g. the conversion to $\rm D_2H^+$ and $\rm D_3^+$).  \hhdp-identified cores which partially overlap with continuum structure, but which present a continuum flux peak outside their boundaries, may belong to this evolutionary stage. Cores 4 and 6 in AG354 are representative of this case (as shown in Appendix \ref{App:contCores}). The fact that they do not present significant increase of the velocity dispersion in the proximity of the continuum peaks suggest that the decrease of \xmol is most likely due either to HD depletion onto dust grains or to conversion into other isotopologues. Finally, cores which are only visible in continuum are the most evolved. The lack of detection of \hhdp in their surroundings suggests that either their temperature is overall higher than $20\rm \, K$, hinting to an already protostellar stage, or that \hhdp is depleted onto dust grains and/or converted into its doubly and triply deuterated forms. \par
In order to test our hypothesis, we have searched for correlations of the suggested evolutionary phases and the cores dynamical properties reported in Table \ref{CoreProp} and \ref{CoreDyn}, such as linewidths, virial masses, and average densities. We did not find any significant trend. Nevertheless, Fig. \ref{Scatt2} shows a decrease of \xmol with an increase of $N(\mathrm{H_2})$, that could resemble a situation where the collapse is more advanced at least in some of the cores. Further observations of tracers of protostellar activity (such as SiO jets or CO outflows) or kinematics tracers could confirm this hypothesis. \par
\begin{figure}[h]
\centering
\includegraphics[width=0.5\textwidth]{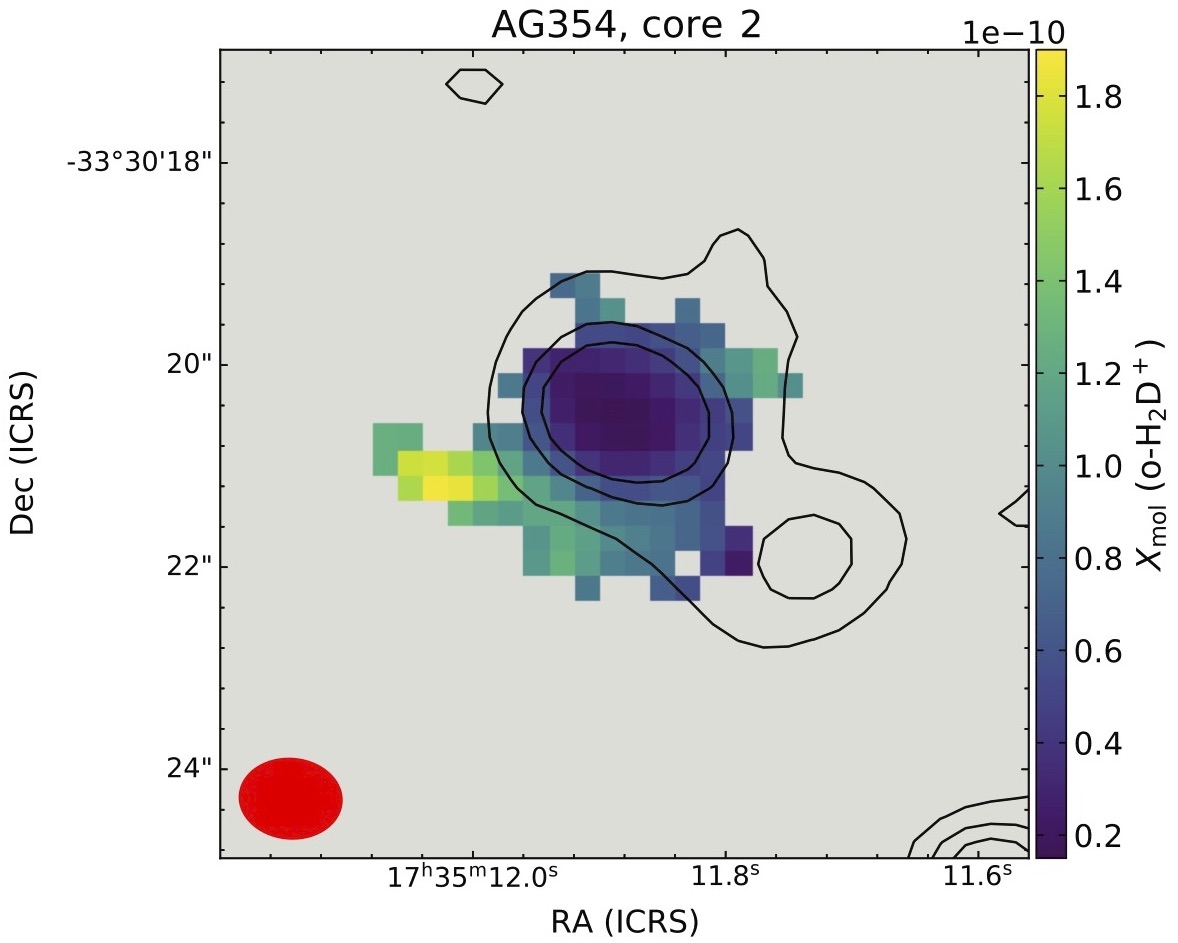}
\caption{The \xmol distribution in core 2 in AG354, the most massive structure identified in this work. \xmol has been computed adopting a constant $T_\mathrm{dust} = 10 \, \rm K$. The contours show the ALMA continuum flux in band 7, at levels $[5, 10, 15]\, \rm mJy \, beam^{-1}$. In the assumption of constant dust temperature, the contours represent also the distribution of $N\rm (H_2)$. \label{PrestCor} }
\end{figure}

According to our analysis, core 3 and 7 (AG351) and core 2 (AG354) represent good candidates to be truly prestellar cores in high-mass clumps, since they overlap significantly with continuum-identified structures with masses $M \ge14 \, \rm M_\odot$. Fig. \ref{PrestCor} shows the distribution of the \ohhdp abundance in core 2 in AG354, the most massive structure we identify, with contours from the continuum (thus representative of the H$_2$ column density, in the assumption of constant dust temperature). It can be seen that the molecular distribution seems anticorrelated with the continuum one: at the dust peak \xmol present the lowest values, whilst the molecular abundance increases moving away from it, in a similar fashion to the general trend shown by the scatterplots in Fig. \ref{Scatt2}. This core shows particularly low linewidths ($\langle \sigma_\mathrm{V} \rangle = 0.20\,$\kms), which become as low as $0.10\,$\kms towards the core centre, where hence the temperature sinks to $\approx 5 \,$K. If we assume this dust temperature, the core mass is $M\rm _{core} =39 \, \rm M_\odot$, significantly more massive than all the other cores studied in this work, and the mass of the associated continuum-identified structure (c-9 in Appendix \ref{App:contCores}) could be as high as $60 \, \rm M_\odot$. Core AG354-2 hence represents an ideal candidate of HMPCs. \par
The evolution here discussed can either be chemical or physical (i.e. a density evolution). The first case holds if the cores are virialised by other sources of pressure other than thermal and turbulent motions, and hence their density is kept approximately constant while the chemistry evolves. In the initial stages, due to the high volume densities and low temperatures, \ohhdp is efficiently formed, and hence the cores show bright molecular emission. As time passes by, the \ohhdp abundance decreases, because the molecule is depleted onto the dust grains and/or converted into other isotopologues, until the \olineh at the continuum peaks is not detectable anymore. \par
In the case the cores are instead subvirial, and hence they are experiencing gravitational contraction, the density is evolving and increasing with time. The evolution of \xmol is however similar, because the high initial abundance is then lowered for instance when the increasing density causes more depletion onto dust grains. In this case, at a certain point the collapse will form a central protostar, whose feedback will reduce the \ohhdp abundance below the detection limit. Lastly, we cannot exclude a combination of the two cases, where cores experience periods of time in equilibrium, chemically evolving at a constant density, followed by contraction phases.

\subsection{Our results in the context of star formation theories\label{Disc:4}}

The analysis of our data shows that the \hhdp-identified cores have masses in the range $0.1-13 \, \rm M_\odot$ depending on the assumed temperature. These values, which are probably accurate within a factor of $\approx 5$ (see Appendix \ref{APEX}), would hence suggest that several of the \hhdp-identified cores are essentially low-mass. The only clear exception is core AG354-2 for which, in the assumption of a dust temperature value $T_\mathrm{dust} = 5\, \rm K$, we estimate $M_\mathrm{core} = 39 \pm 13 \, \rm M_\odot$. We highlight that in order to form a $8\, \rm M_\odot$ star, a core of at least $30 \, \rm M_\odot$ is needed, assuming a star formation efficiency of 30\%.   \par
Furthermore, both at core and clump scales the analysed sources show sub-virial conditions, in contrast with the predictions of turbulent-core accretion models of HMPCs. In our sample of \hhdp-identified cores, 80\% of them are consistent with $\alpha_\mathrm{vir} \lesssim 1$, regardless of the assumed temperature. The virial parameter computed in the previous section however takes into account only the kinetic energy content ($T$) and the gravitational energy ($U$), whilst in a more complete fashion one has to take into account also the magnetic energy term, which in case of a homogeneous and spherical core threaded by a uniform magnetic field $B$ reads:
\begin{equation}
\Omega_B= \frac{4 \pi}{3} R_\mathrm{eff}^3 \times \frac{1}{8 \pi }B^2 \; ,
\end{equation} 
where the equation is expressed in cgs units. We can determine the strength of the magnetic field needed to halt the gravitational collapse by imposing the virialisation: $U+2T+\Omega_B = 0$. In our sample of cores, the estimated $B$ values range from hundreds of $\mu \rm G$ to several $ \rm mG$. Direct magnetic field measurements via the Zeeman effect are rare, but we note that in \cite{Crutcher19} at densities of $n \approx 10^5 -10^6 \, \rm cm^{-3}$ the authors report values of $B \approx \text{a few hundreds}\,\rm \mu G$, but never higher than $1\, \rm mG$. However, analysing the  polarised dust thermal emission, several authors found magnetic fields of the order of the milliGauss in massive star-forming regions \citep{Girart13, Zhang14, Liu20}. In conclusion, it is plausible that magnetic fields provides the extra support needed to virialise the cores, even though for those with very low virial ratios ($\alpha_\mathrm{vir}<0.3$), magnetic strengths of several $\rm mG$ would be needed. We highlight the need of further observations to investigate the level of magnetisation of the clumps, and their large-scale kinematics. \par
We have also estimated the Jeans masses for the \hhdp-identified cores. For volume densities in the range $n = 10^{6} - 10^7 \, \rm cm^{-3}$ and temperatures of $5-10\, \rm K$, the Jeans mass is $M_\mathrm{Jeans} = 0.03-0.3 \, \rm M_\odot$. Hence, almost the totality of the \hhdp-identified cores contains several tens, up to hundreds, of Jeans masses. This will cause a large degree of further fragmentation in the more massive cores, unless magnetic pressure or a high-level of turbulence are assumed. The analysis we have performed on the linewidths show that the motions of the gas traced by \olineh is at most slightly supersonic ($\sigma_\mathrm{V,NT}/ c_\mathrm{s} < 2$ in all cases), similarly to what is found at the clump level (see \citealt{Sabatini20}), pointing towards the necessity of magnetic fields to prevent fragmentation. \par
If we exclude that the mass values derived from the ALMA observations are significantly underestimated, it looks reasonable to evaluate two alternative possibilities: \textit{i)} these cores are going to form low- to intermediate-mass stars if no further accretion is considered, with perhaps the exception of core AG354-2, or \textit{ii)} these cores can be clump-fed and eventually form high-mass stars as in the competitive accretion model \citep{Bonnell06}, growing during the low-mass protostellar stage similarly to what also proposed by \citet{Tige17,Motte18}. We have highlighted how AG351 and AG354 sits in the low-end of the clump masses in the ATLASGAL sample.  \cite{Kauffmann10} investigated empirically the threshold for high-mass star formation in galactic IRDCs, and according to their Eq. 1 our clumps are above the mass limit, and in principle they can form high-mass stars. On the other hand, \cite{Sanhueza17} computed the minimum clump mass needed to form at least one massive star ($M> 8 \, \rm M_\odot$), assuming a Kroupa initial mass function (IMF), and a star formation efficiency of 30\%, which is $M_\mathrm{clump} ^\mathrm{lim} = 260 \rm \, M_\odot$. This threshold is however strongly dependent on the assumed IMF shape and star-formation efficiency, and it is hence very uncertain. We therefore cannot exclude that more massive clumps, representative of the high-mass clumps population ($M \approx 500  \rm \, M_\odot$), can host several high-mass prestellar cores. Further studies focused not only on the continuum emission but also on \hhdp on more massive clumps are then needed to assess the existence of massive prestellar cores and disentangle between the different theories.
\par
Up to now, in literature the best candidates of HMPCs have been the C1S core dentified by \cite{Tan13} ($M = 10-50 \rm \, M_\odot$ depending on the assumed temperature), the one by \cite{Cyganowski14} ($30 \rm \, M_\odot$) and the core W43-MM1\#6 studied by \cite{Nony18}. In the first study, the authors reported high abundances of \nndp and in a follow-up paper a non-detection of \ohhdp \citep{Kong16}. Based on the results from \cite{Giannetti19} (see Sect. \ref{Introduction} for more details) the C1S core could hence already host an embedded protostar in its very initial stages, or it could be perturbed by the activity of the two nearby, protostellar cores seen by \cite{Tan16}. The source investigated by \cite{Cyganowski14}, on the other hand, has been identified only in continuum, which prevents from a conclusive assessment of its evolutionary stage. Core 6 in W43-MM1, with $M = 60 \rm \, M_\odot$ and no outflow detected, remain a good candidate to be an isolated and massive HMPCs. Future observations targeting molecular tracers of dense and cold gas are needed to validate this hypothesis. Recent studies (see e.g \citealt{Sanhueza19, Contreras18}) pursued with high-resolution ALMA observations revealed a large population of low-mass cores in high-mass clumps, showing subvirial dynamical states, in agreement with what we report in this study. 
\par
The main advantage of this work is the detection of \ohhdp, which traces the very early stages, and we can then state these are possibly the first prestellar cores identified unambiguously in high-mass star-forming clumps. Further observations focusing on the magnetic properties of the observed cores will provide us with the detailed analysis of the magnetic fields needed to assess conclusively the their dynamical states. In addition, numerical studies including chemistry are needed to shed light on the different proposed theories and disentangle among the different physical processes that can affect the star-formation process. Examples of these works, including detailed deuteration chemistry, have recently started to be developed \citep{Goodson16, Kortgen18, Bovino19, Hsu21} and a proper comparison of these simulations with observations is a viable way to find an answer to this longstanding problem.

\section{Summary and Conclusions\label{Summary}}
In this work, we report ALMA observations in band 7 at a resolution of $\approx 1''$ in two quiescent, intermediate-to-high mass clumps. For the first time, we report the detection of \olineh in this kind of sources with interferometric observations. Our molecular line data show that \ohhdp is very extended, and its distribution does not correlate with the one of the dust thermal emission in the same ALMA band ($\lambda = 0.8\, \rm mm$). \par
Using the algorithm \textsc{scimes}, we have identified 16 cores in the \ohhdp datacubes. We have fitted their spectra pixel-per-pixel using a Bayesian approach implemented in the code \textsc{MCWeeds}, in order to derive the line velocity dispersion, the molecular column density, and the centroid velocity map. The first important conclusion is the detection of narrow \olineh lines, with linewidths lower than their thermal broadening at $10\, \rm K$. This indicates that this species traces very cold and quiescent region in the analysed sources. \par
We have investigated the general lack of correlation between the dust continuum peaks and the molecular line peaks, which has profound implications. We suggest that this is due to a possible physical and/or chemical evolution. In the initial stages \ohhdp is quite abundant, and tracing the high-density gas. The presence of cores bright in the molecular emission, but lacking a continuum peak nearby, can be due to the fact that the ALMA band 7 observations are not able to trace the peaked dust distribution due to temperature effects (see Sec. \ref{Disc:3} for more details). Later on, as the gas becomes denser, or as the chemistry evolves, the \xmol starts to decrease, but the molecule is still detectable. Eventually, the \hhdp abundance drops, due either to depletion at very high densities, or to protostellar feedback effects. \par
Our results highlight how the continuum emission alone is generally not a good probe of prestellar gas. Bright cores in continuum flux which do not show significant emission in cold-gas tracers (such as \hhdp) are likely in a more evolved, possibly protostellar stage. This suggests that particular care must be taken when doing surveys of  ``prestellar'' cores seen only in continuum, especially when only one frequency band is available, since these data do not allow to determine the temperature, the evolutionary stage, and the kinematic properties of the sources. Complementary molecular data should be used to distinguish pre- and proto-stellar objects. In this context, we find that \ohhdp represents a good tracer of cold gas at densities of $n \approx 10^{6} \, \rm cm^{-3}$. This gas is in a prestellar phase, in the sense that it has not be influenced by protostellar activity, and ---being cold and dense--- it has the potential to form new protostars; its evolution, however, is determined by its dynamic state. \par
At densities higher than $10^{6} \, \rm cm^{-3}$ even \ohhdp could be not ideal anymore, as shown by its abundance drop as a function of $N(\rm H_2)$, due to opacity effects, depletion, or a combination of both. In those physical conditions, $\rm D_2H^+$ and $\rm D_3^+$ represent probably the only good tracer available. Whilst the latter is not observable, the former could represent, in combination with \ohhdp,  the best choice to investigate the more evolved and denser regions. \par
Most of the \hhdp-identified cores are less massive that $10 \, \rm M_\odot$, even in the assumption of low dust temperatures ($5\, \rm K$). It is important to highlight, however, that the ALMA observations could be filtering out the most extended core envelopes, hence leading to underestimation of their total masses. Furthermore, we cannot exclude that they are still in the process of accreting material from the parental clump. On the other hand our data could support a scenario in which high-mass clumps fragment in a population of low-to-intermediate cores, which can continue the accretion to larger masses during the later, protostellar phase. Two of the cores contains peaks of both molecular and continuum emission, which allows us to reliably estimate their total masses. Core AG354-2 shows particularly narrow linewidths, consistent with temperature values as low as $5\, \rm K$. With this dust temperature, we estimate a core mass of $39 \, \rm M_\odot$, which could represent a lower limit for the aforementioned reasons. Furthermore, it is associated with a continuum-identified core, which contains several tens of solar masses. To our knowledge it hence represents an ideal candidate of HMPCs, but further investigation is needed to better constrain its temperature and thus mass. \par
We have investigated the dynamic state of the \hhdp-identified cores by means of the virial analysis. Most of the cores appear subvirial if we take into account the gravitational energy and the kinetic energy only. We however do not have information on the magnetic properties of the sources. We estimate that magnetic field strengths of the order of several hundreds of microGauss or a few milliGauss are needed to virialise the cores. Further observations aimed to recover the magnetic field properties are needed to make a definite conclusion on the core dynamic states. \par
As future perspective, we plan to recover the missing flux in the large scale emission of \olineh (see Appendix \ref{APEX}). This will allow us to use this transition to investigate the kinematics of the gas at the clump scales, which in turn will provide key information about the cores' dynamics. Furthermore, ALMA observations at a similar spatial resolution of molecules which are good tracers of the protostellar activity (SiO, CO,...) will help us disentangle unambiguously prestellar cores from protostellar ones.

 \begin{acknowledgements}
   {We thank the anonymous referee, for her/his suggestions to improve the manuscript.}
This paper makes use of the following ALMA data: ADS/JAO.ALMA\#2018.1.00331.S. ALMA is a partnership of ESO (representing its member states), NSF (USA) and NINS (Japan), together with NRC (Canada), MOST and ASIAA (Taiwan), and KASI (Republic of Korea), in cooperation with the Republic of Chile. The Joint ALMA Observatory is operated by ESO, AUI/NRAO and NAOJ. In addition, publications from NA authors must include the standard NRAO acknowledgement: The National Radio Astronomy Observatory is a facility of the National Science Foundation operated under cooperative agreement by Associated Universities, Inc. This research made use of \textsc{scimes}, a Python package to find relevant structures into dendrograms of molecular gas emission using the spectral clustering approach. ER is thankful to Dominique M. Segura-Cox, for the help provided in the data reduction. SB acknowledges the BASAL Centro de Astrofisica y Tecnologias Afines (CATA) AFB-17002. The data analysis has been performed with resources provided by the KULTRUN Astronomy Hybrid Cluster at Universidad de Concepci\'on. DC acknowledges support by the \emph{Deut\-sche For\-schungs\-ge\-mein\-schaft, DFG\/} project number SFB956A.
\end{acknowledgements}
 

 \appendix
 
 \section{ALMA and APEX comparison\label{APEX} }
\begin{figure}[!h]
\centering
\includegraphics[width=0.48\textwidth]{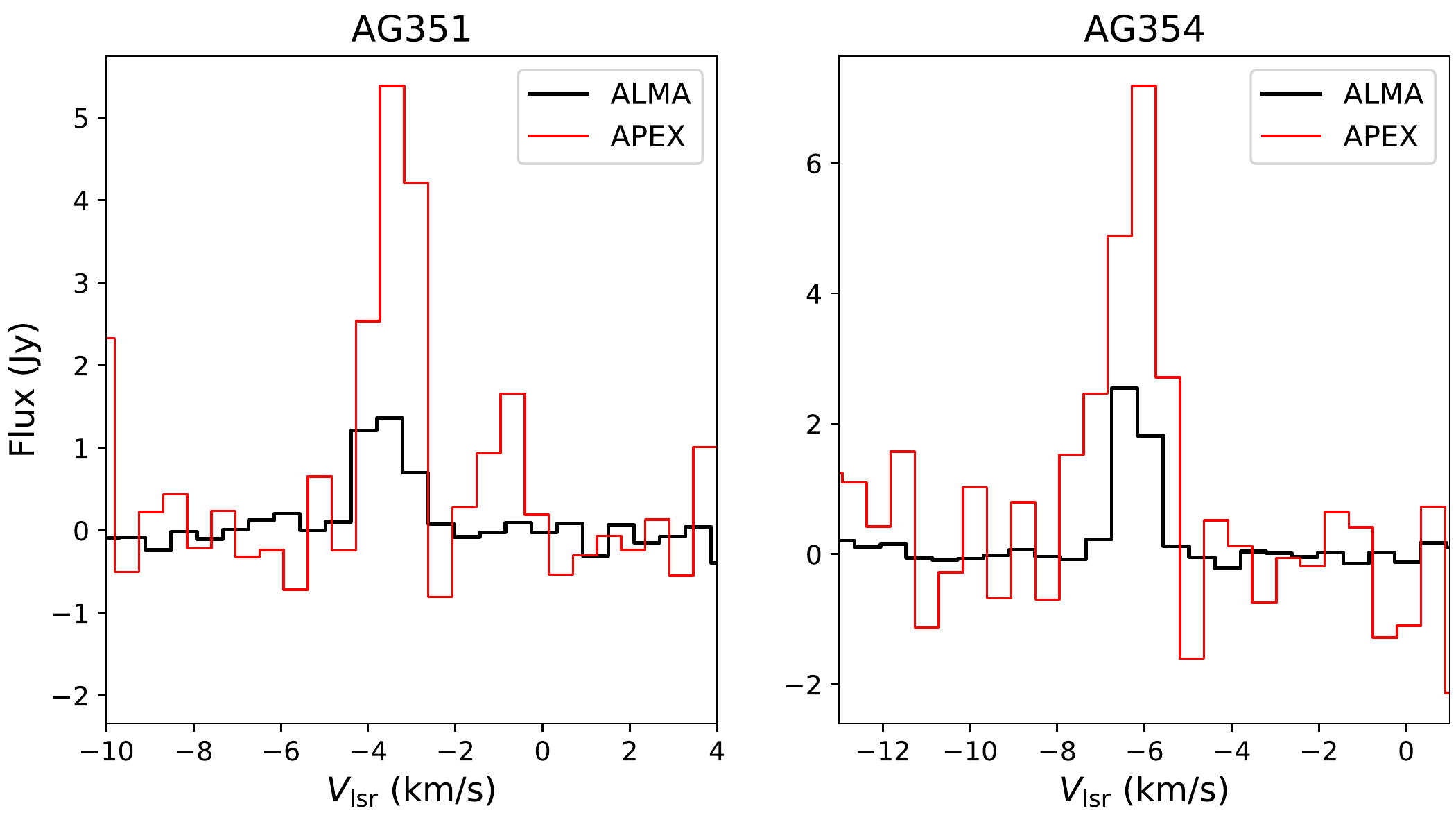}
\caption{Comparison of the \olineh spectra observed by APEX (red) and ALMA (black) in AG351 and AG354, from left to right. For a fair comparison, the APEX spectra from \cite{Sabatini20} have been converted in flux units using the APEX gain ($40 \, \rm Jy \, K^{-1}$). The ALMA spectra have been computed integrating the signal over an area corresponding to the APEX beam size ($16.8''$), and they have been smoothed to the APEX spectral resolution. \label{AlmaApex}}
\end{figure}

In Fig. \ref{AlmaApex} we show the comparison of the \olineh spectra obtained with APEX (from \citealt{Sabatini20}) and with ALMA (this work) in AG351 and AG354. In order to compare the two datasets, we have converted the temperature scale of the APEX observations in flux, using the gain $G = 40 \, \rm Jy/K$ (from the APEX telescope efficiency webpage\footnote{\url{http://www.apex-telescope.org/telescope/efficiency/}}). The ALMA data have been integrated over an area equal to the APEX beam size, and then convolved to the same velocity resolution ($\approx 0.55\,$\kms). \par
The ALMA spectra in both sources present a peak flux of $\approx 1/3-1/5$ of the APEX peak. This is due to the so-called missing-flux problem, which is caused by the fact that the molecular emission is diffuse over scales larger than the maximum-recovable-scale $\theta_\mathrm{MRS} \approx 20''$. As a consequence, a significant amount of the emission is filtered out. However, it is important to notice that the ALMA integrated spectra do not present anomalous line shapes, and their centroid velocity is consistent with the one from the single-dish observations.   {Furthermore the core apparent sizes are smaller than the maximum recoverable scale $\theta_\mathrm{MRS} \approx 20''$. All this considered, we expect} that the missing-flux problem is not affecting severely the line profiles in the cores, and hence that the kinematics parameters (\sigmav, \vlsr) obtained from the spectral fitting are reliable. We however highlight that future observations recovering the zero-spacing flux are needed (see e.g. \citealt{Henshaw14, Sokolov18}), if we want to use the ALMA \olineh data to discuss the large scale kinematics of the clumps.
\par
Concerning the continuum data, we can estimate the fraction of flux filtered out by the interferometer comparing the ALMA and the APEX observations at $870 \, \rm \mu m$ of the clumps from the ATLASGAL project. Integrating over an area equal to the ALMA FoV, we obtain $S_{870\, \rm \mu m} = 2.6 \rm \, Jy$ (AG351) and  $S_{870\, \rm \mu m} = 1.8 \, \rm Jy$ (AG354), whilst the total flux seen in the ALMA data is  $S_{0.8\, \rm m m} = \rm 1.0 \, Jy$ (AG351)  and $S_{0.8\, \rm m m} = 0.93 \, \rm Jy$ (AG354). This is just a rough comparison, since we are not taking into account the $\approx 10$\% difference in the observed wavelength, but it shows that the interferometer is filtering out $\approx 50-60$\% of the emission. \par 
Despite most of the missing flux comes from the large clump scales, we cannot exclude that part of the cores envelopes are also filtered out, hence leading to mass underestimation, as observed in the nearby low-mass prestellar core L1544 by \cite{Caselli19}. It is however not straightforward to determine by which factor the masses are possibly underestimated, since the APEX single-dish observations themselves could underestimate the flux due to filtering of the large scale emission. Hence also the clump total masses could be underestimate, and the total mass budget available for the cores could be higher than $\approx 150\, \rm M_\odot $. In a conservative approach, we estimate that the mass values could be underestimated of a factor $2-5$. 

 \section{Continuum-identified cores\label{App:contCores}}

   \begin{figure*}[h]
\centering
\includegraphics[width=\textwidth]{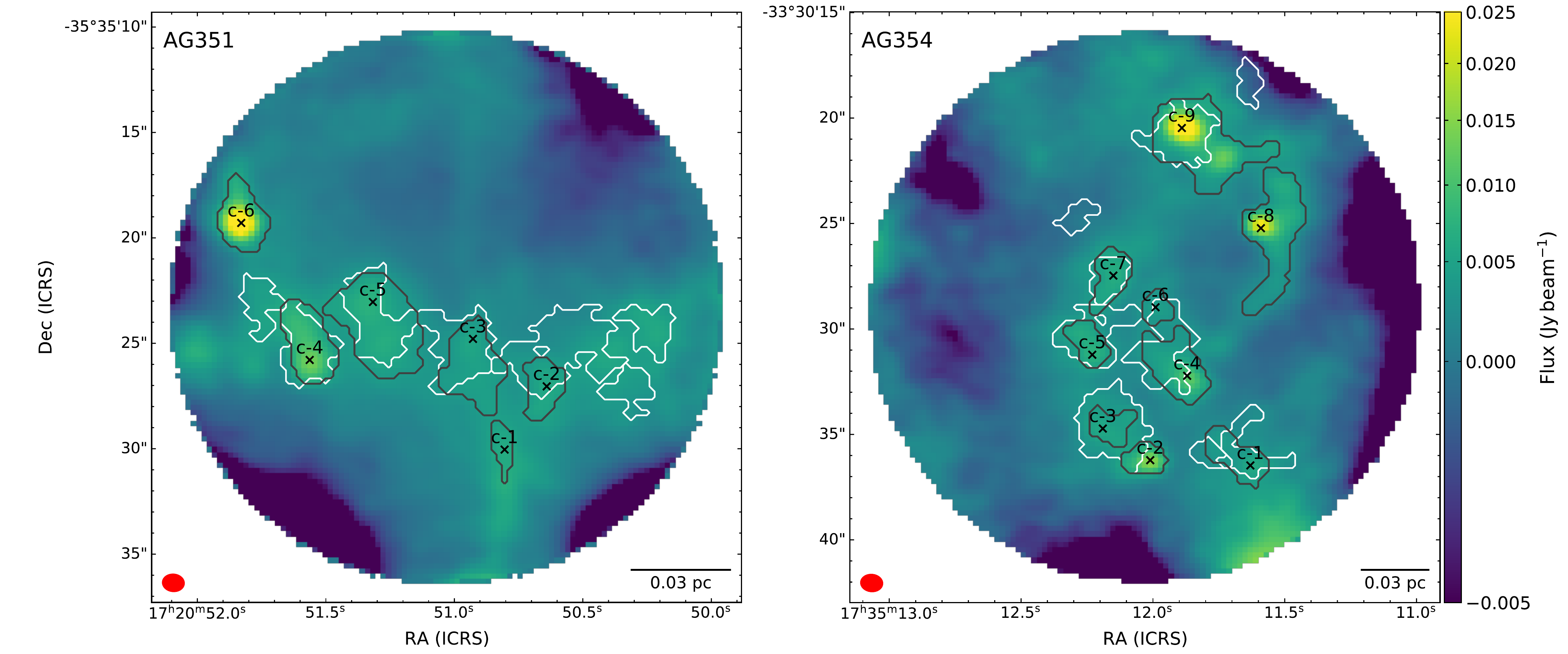}
\caption{The background image show the dust thermal emission in band 7, with overlaid the continuum-identified structures as grey contours. The black crosses show the position of the flux peak within each core. The white contours show the \hhdp-identified cores resulting from the \textsc{scimes} analysis. \label{ContCores}}
\end{figure*}

  \begin{table*}[h]
   {
 \renewcommand{\arraystretch}{1.4}
\centering
\caption{Properties of the cores identified in continuum emission. The masses are estimated at three different temperatures. In the last two columns we report the properties of the overlapping \hhdp-identified cores.}
\label{Table:contCores}
\begin{tabular}{ccccccc}
\hline
Core id\tablefootmark{a}      &    $R_\mathrm{eff}$ & \multicolumn{3}{c}{$M_\mathrm{core}/ \rm M_\odot$}   & Corresponding & Fraction of \\
	& 	$10^3$AU	& $T_\mathrm{dust} = 5 \, \rm K $& $T_\mathrm{dust} = 10 \, \rm K $  & $T_\mathrm{dust} = 20\, \rm K $ & \hhdp core\tablefootmark{b}       & overlap\tablefootmark{c} \\
\hline \hline
\multicolumn{7}{c}{AG351}                                                                \\
c-1 & 1.0 &  $2.7\pm 0.9$ &   $0.39\pm 0.13$    &   $0.11\pm 0.04$ & -&	0\%		\\ 
c-2 & 1.5 &  $5.1\pm 1.7$ &   $0.7\pm 0.2$    &   $0.21\pm 0.07$ & 5& 	59\%			\\  
c-3 & 2.2 &  $9\pm 3$ &   $1.4\pm 0.5$    &   $0.40\pm 0.13$  & 6&	60\%	\\  
c-4 & 2.0 &  $14\pm 5$ &   $2.0\pm 0.7$    &   $0.59\pm 0.19$ & 3& 83\% \\  
c-5 & 2.9 &  $19\pm 6$ &   $2.7\pm 0.9$    &   $0.8\pm 0.3$ & 7&		56\% \\  
c-6 & 1.8 &  $18\pm 6$ &   $2.6\pm 0.9$    &   $0.8\pm 0.3$ & - &	0\%		\\  
\hline
\multicolumn{7}{c}{AG354}                                                                                                                        \\
c-1 & 2.4 &  $10\pm 3$ &   $1.5\pm 0.5$    &   $0.42\pm 0.14$ &3 & 57\% \\  
c-2 & 1.7 &  $11\pm 4$ &   $1.6\pm 0.5$    &   $0.47\pm 0.16$  &4&		44\%	\\  
c-3 & 1.8 &  $7\pm 2$ &   $1.1\pm 0.4$    &   $0.31\pm 0.10$ &4 &		100\% \\ 
c-4 & 2.9 &  $20\pm 7$ &   $2.9\pm 0.9$    &   $0.8\pm 0.3$ &6& 		56\%\\  
c-5 & 1.9 &  $9\pm 3$ &   $1.3\pm 0.4$    &   $0.39\pm 0.13$& 8&		98\% \\  
c-6 & 1.5 &  $3.6\pm 1.2$ &   $0.52\pm 0.17$    &   $0.15\pm 0.05$ &6 &	60\%\\  
c-7 & 2.1 &  $11\pm 4$ &   $1.5\pm 0.5$    &   $0.45\pm 0.15$ & 7&	70\%	\\  
c-8 & 3.6 &  $34\pm 11$ &   $5.0\pm 1.6$    &   $1.4\pm 0.5$& - &		0\%\\  
c-9 & 4.1 &  $60\pm 20$ &   $9\pm 3$    &   $2.7\pm 0.9$& 2 &			45\%\\  
\hline                
\end{tabular}
\tablefoot{
\tablefoottext{a}{The cores are labeled as c-N, to avoid confusion with the \hhdp-identified ones.} \\
\tablefoottext{b}{Label of the \hhdp-identified structure (according to Table \ref{CoreProp}) which presents the largest overlap with the selected continuum-identified core.}\\
\tablefoottext{c}{Fraction of the area of continuum-identified core which overlaps with the corresponding \hhdp-identified one reported in the 6th column, if applicable, expressed as percentage.}

} 

}
\end{table*}
 In the main text we have focused on the properties of the \hhdp-identified cores. However, it is worth looking also to the continuum-identified structures, which is the purpose of this Appendix. \textsc{scimes} is developed to work in position-position-velocity space, and hence it is suitable to analyse molecular line data only. We have hence used the python package \textsc{astrodendro} (\url{http://www.dendrograms.org/}), on which \textsc{scimes} is also built, to identify structures in the $0.8 \rm \, mm$ continuum maps. After a few tests followed by visual inspection of the results, we have selected the dendrogram parameters as it follows: $ {min_\mathrm{val}} = 3\sigma$, $\Delta_\mathrm{min} = 1\sigma$, and $min_\mathrm{area} = 2 \, \rm beam$ (i.e. we exclude structures smaller than two beams), a choice consistent with similar works (see e.g. \citealt{Barnes21}). As previously stated, dendrogram analysis algorithms perform better on constant-noise maps, and hence we input $1\sigma = 0.5 \, \rm mJy \, beam^{-1}$ as the $rms$ level of the continuum maps before the correction for the primary beam response.  \par
We find in total 15 cores, which are shown in Fig. \ref{ContCores}, together with the \hhdp cores identified in Sect. \ref{SCIMES}. As previously discussed, there is no one-to-one correspondence in the identified structures. Some of the brightest figures seen in dust thermal emission do not correspond to cores visible in \ohhdp, and vice versa. Several structures overlap, but it is interesting to notice that for some of them (c-2 and c-4 in AG354) the continuum peak is found at the border, or just outside, of the \hhdp-identified cores. In Table \ref{Table:contCores} we report the label of the corresponding \hhdp-identified core that overlaps the most with each of the continuum-identified ones, together with the fraction of overlap. One can see that several structures overlap only partially. For nine of the continuum-identified cores, the overlapping fraction is in the range $0-70$\%. 
\par
Similarly to what done for the \hhdp-identified cores, we have computed the sizes in term of effective radii and the masses of the cores identified in continuum emission. As extensively discussed in Sects. \ref{Disc:1} and \ref{Disc:2}, the assumption of constant dust temperature $T_\mathrm{dust} = 10\, \rm K$ might not be appropriate. We have hence computed \mcore via Eq. \eqref{Mdust} at three different temperatures, i.e. $5, 10, \text{ and } 20\, \rm K$, which cover the range of $T_\text{\hhdp}$ values. Similarly to Table \ref{CoreDyn}, we estimate a 33\% relative error on the masses. We highlight again that the lack of total-power data (which is however not offered in continuum observations with ALMA) may cause a partial underestimation of the total core masses, in the hypothesis that part of the extended emission is filtered out. The resulting parameters are summarised in Table \ref{Table:contCores}.\par
The continuum-core sizes are similar to those of the \hhdp-identified ones, and present an average effective radius of $2.2\, \times 10^3 \rm AU$. In general, the cores present low masses, unless low dust temperature values are assumed, which however are supported by our \ohhdp analysis. In AG351, three cores are expected to be more massive than $10\, M_\odot$ in the assumption of $T_\mathrm{dust} = 5 \, \rm K $. Core c-4 and core c-5 were identified also in the \ohhdp data. Core c-6, instead, do not present bright emission in the molecular line, which may be due either to high depletion of \hhdp, or to the presence of protostellar feedback from an unidentified protostellar object, as discussed in Sect \ref{Disc:3}.\par
Continuum cores in AG354 are on average more massive than in AG351. Six out of nine cores present in fact $M_\mathrm{core} \ge 10 \, M_\odot$ at $5 \, \rm K$. Core c-9, which correspond to the \hhdp-identified core 2, appears as massive as $60\, \rm M_\odot$. It is worth noticing that core c-9 contains two emission peaks, separated by $2.5''$, the fainter of which has a peak flux approximately one-third of the brighter one. In the dendrogram analysis they however are never separated, despite the choice of input parameters, most likely due to the fact that they lay closer than  three ALMA beams within each other. Core c-8 is in a similar situation as core c-6 in AG351, since it does not correspond to bright molecular emission, and hence could be in a later evolutionary stage.\par
Overall, the dendrogram analysis confirms our findings: most cores are essentially low-mass, even though we stress again that we may be underestimating masses from the continuum emission due to the lack of zero-spacing observations. Furthermore, the correlation between continuum- and \hhdp-identified structures is poor, since the majority of cores either do not overlap completely in the two datasets, or they do for less than 70\% of their projected area. There are however a few exceptions, the most important of which is represented by core c-9, which, with an estimated mass budget of several tens of solar masses and detectable \ohhdp emission, represents an ideal HMPC candidate.

\newpage

\section{Parameter maps of individual cores \label{App:allcores}}

 Figs. \ref{Panel1} to \ref{Panel3} show the maps of the best-fit parameters obtained with \textsc{MCWeeds} in each core.

 \begin{figure*}[h]
\centering
\includegraphics[width=0.9\textwidth]{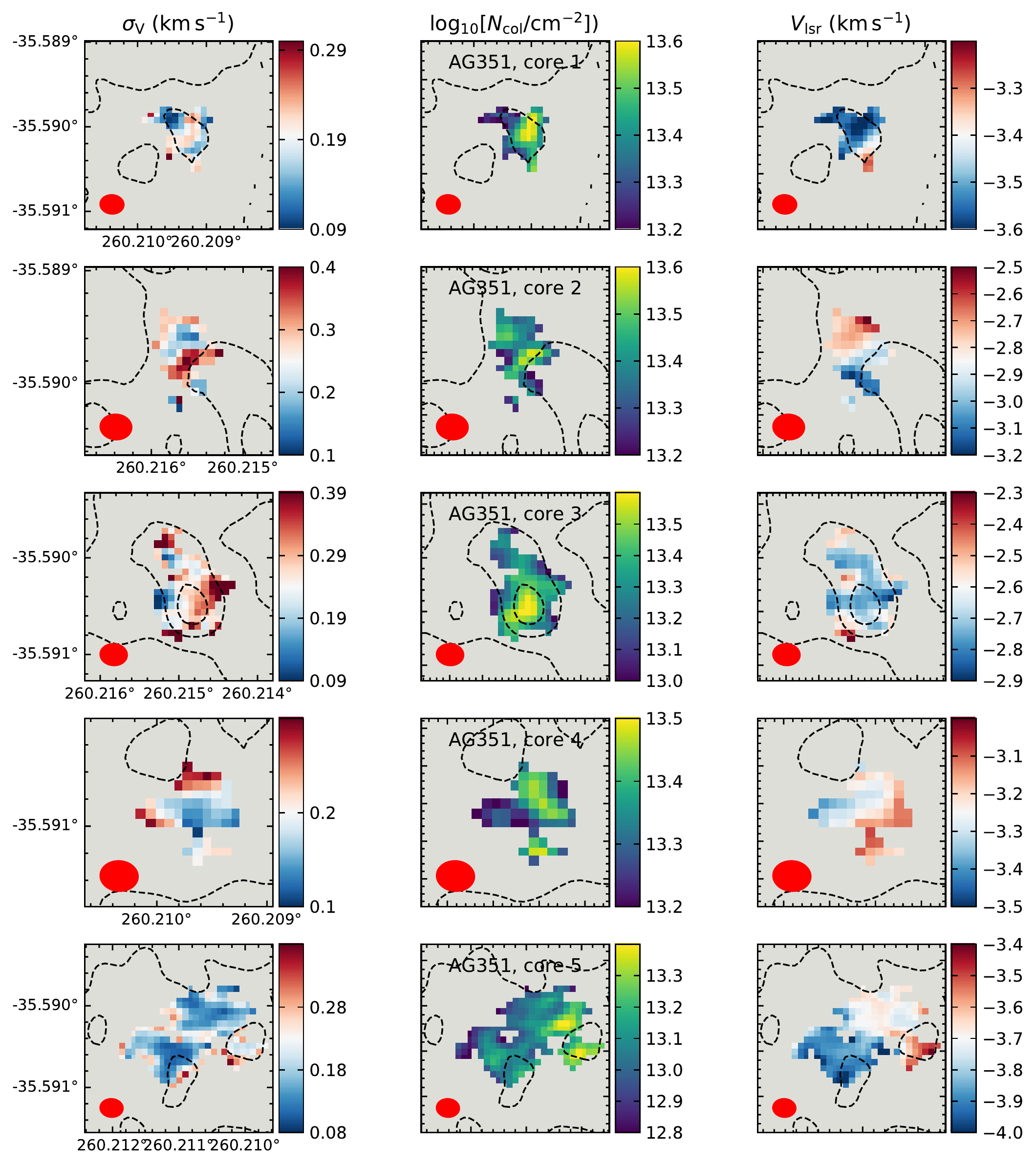}
\caption{\textit{Left panels:} maps of \sigmav, in \kms units; \textit{central panels:} maps of the molecular column density of o-\hhdp, in unit of $\log_{10} \rm (cm^{-2})$; \textit{right panels:} maps of \vlsr, in \kms units. Each row shows a different core, labelled at the top of the central panel. The beam size of the \hhdp observations is shown in the bottom-right corners. The coordinate are shown in the ICRS system. The dashed contours show the continuum emission at levels $[1, 5, 10, 15, 20]\, \rm mJy\, beam^{-1}$\label{Panel1}}
\end{figure*}

  \begin{figure*}[h]
\centering
\includegraphics[width=0.9\textwidth]{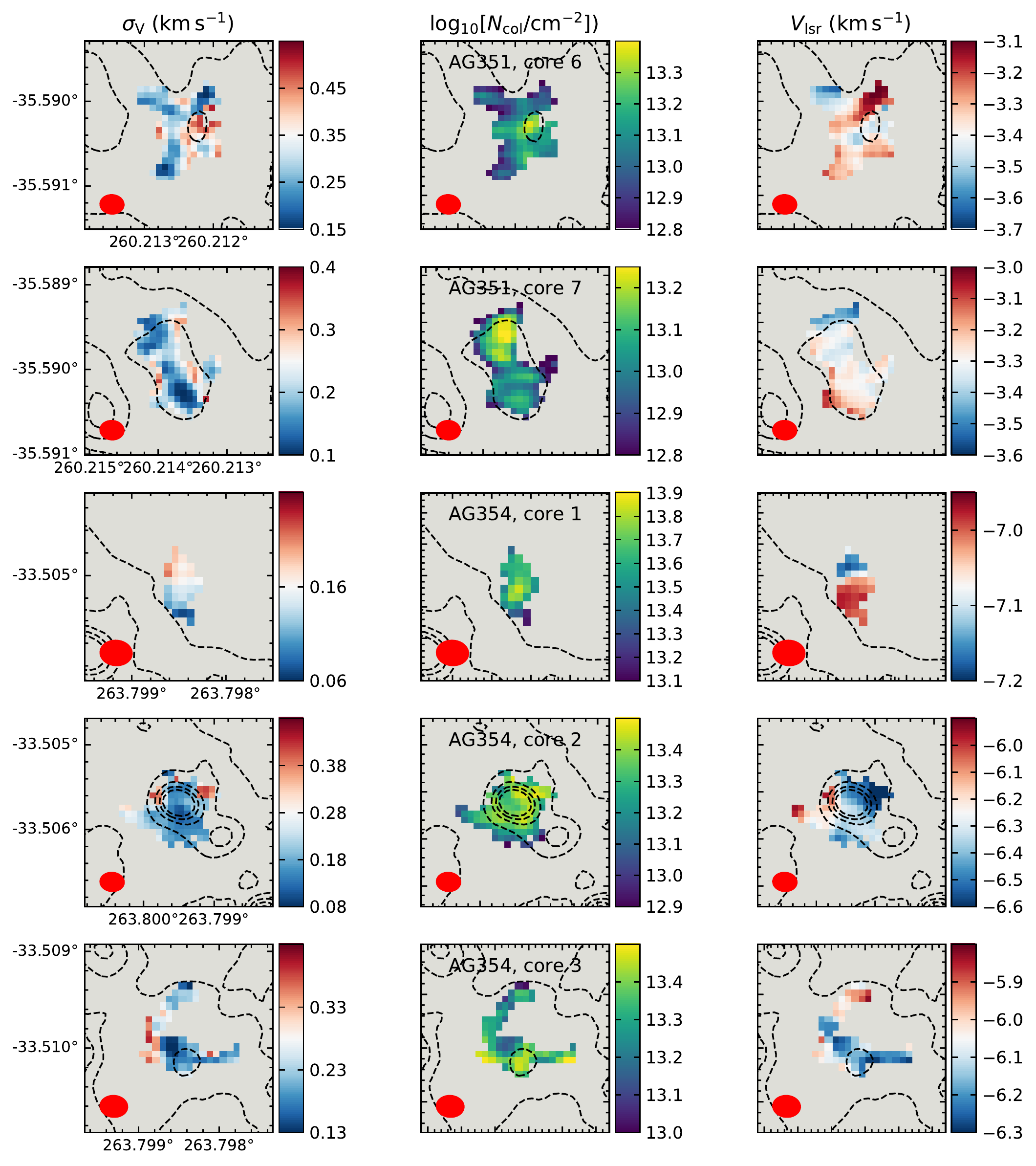}
\caption{Continuation of Fig. \ref{Panel1}. \label{Panel2}}
\end{figure*}

 \begin{figure*}[h]
\centering
\includegraphics[width=0.9\textwidth]{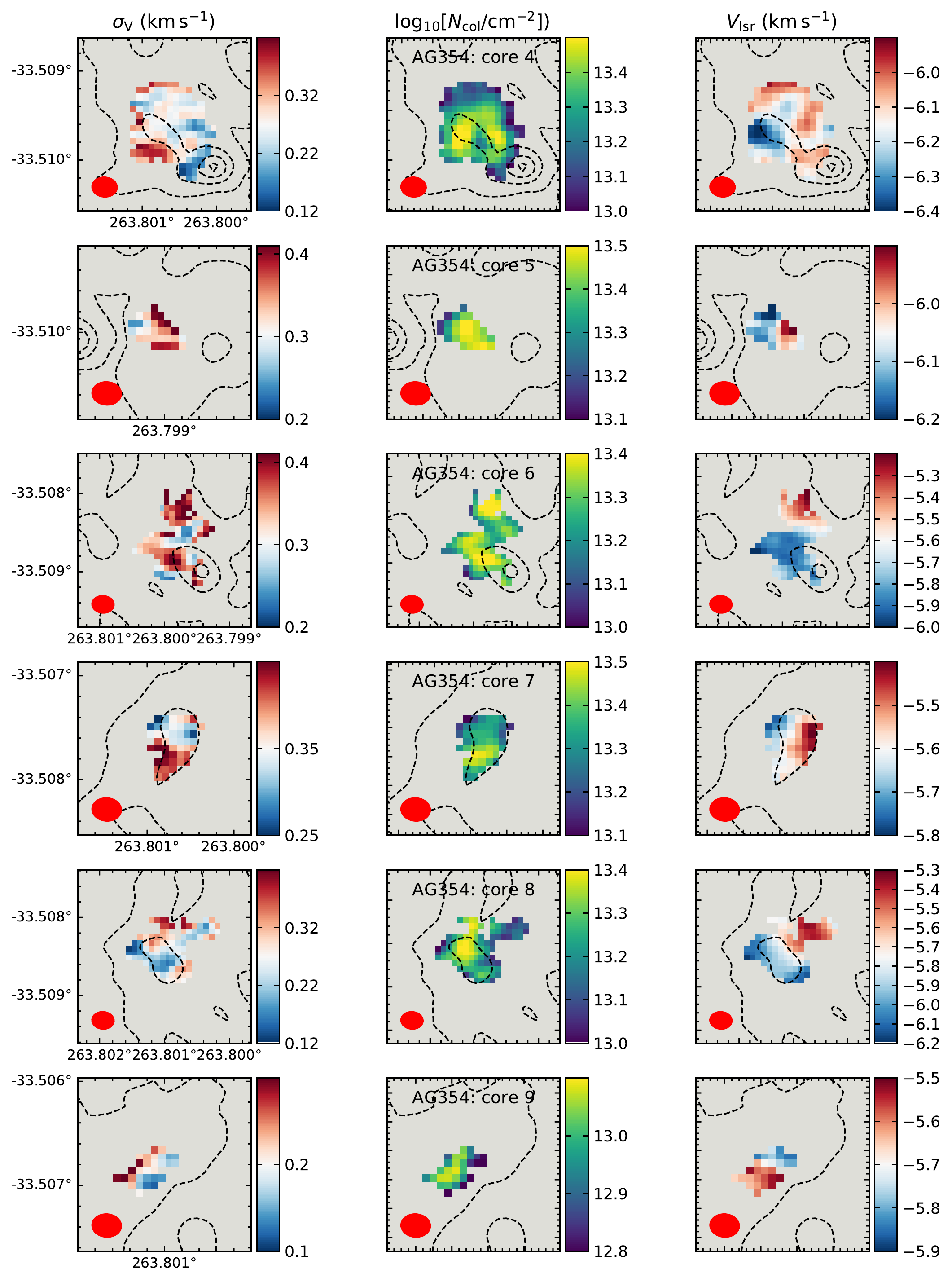}
\caption{Continuation of Fig. \ref{Panel1}. \label{Panel3}}
\end{figure*}

 \end{document}